\documentclass[final,5p,times,twocolumn]{elsarticle}

\usepackage{amssymb}

\usepackage{amsmath}

\usepackage[per-mode=symbol]{siunitx}
\DeclareSIUnit\mwe{m.w.e.}

\usepackage{xspace}
\newcommand{\Teff}{\ensuremath{T_{\rm eff}}\xspace}
\newcommand{\Aeff}{\ensuremath{A_{\rm eff}}\xspace}
\newcommand{\Emu}{\ensuremath{E_{\mu}}\xspace}
\newcommand{\Emumin}{\ensuremath{E_{\mu}^\mathrm{min}}\xspace}
\newcommand{\aT}{\ensuremath{\alpha_{T}}\xspace}
\def\dd{\mathrm{d}}
\newcommand{\rKpi}{\ensuremath{r_{K/\pi}}\xspace}
\newcommand{\Pint}{\ensuremath{P_\mathrm{int}}\xspace}

\newcommand{\refeq}[1]{Eq.~(\ref{#1})}
\newcommand{\refeqs}[2]{Eqs.~(\ref{#1})~and~(\ref{#2})}
\newcommand{\refeqss}[3]{Eqs.~(\ref{#1}), (\ref{#2})~and~(\ref{#3})}
\newcommand{\reffig}[1]{Fig.~\ref{#1}}
\newcommand{\reffigs}[2]{Figs.~\ref{#1}~and~\ref{#2}}

\newcommand{\refsec}[1]{Section~\ref{#1}}
\newcommand{\refsecs}[2]{Sections~\ref{#1}~and~\ref{#2}}
\newcommand{\refsecss}[3]{Sections~\ref{#1}, \ref{#2}~and~\ref{#3}}

\newcommand{\reftab}[1]{Table~\ref{#1}}

\newcommand{\refref}[1]{Ref.~\cite{#1}}

\usepackage{hyperref}

\journal{Astroparticle Physics}

\begin{document}

\begin{frontmatter}



\title{Atmospheric muons and their variations with temperature}


\author[1]{S.~Verpoest\corref{cor1}}
\affiliation[1]{organization={Bartol Research Institute and Dept. of Physics and Astronomy,
University of Delaware},
            city={Newark},
            postcode={19716}, 
            state={DE},
            country={USA}}
\cortext[cor1]{Corresponding author}
\ead{verpoest@udel.edu}

\author[2,3]{D.~Soldin}
\affiliation[2]{organization={Department of Physics and Astronomy, University of Utah},
            city={Salt Lake City},
            postcode={84112}, 
            state={UT},
            country={USA}}
\affiliation[3]{organization={Karlsruhe Institute of Technology, Institute of Experimental Particle Physics},
            city={Karlsruhe},
            postcode={D-76021}, 
            country={Germany}}
\author[4]{P.~Desiati}
\affiliation[4]{organization={Wisconsin IceCube Particle Astrophysics Center, University of Wisconsin--Madison},
            city={Madison},
            postcode={53703}, 
            state={WI},
            country={USA}}

\begin{abstract}
Seasonal variations of atmospheric muons are traditionally interpreted in terms
of an effective temperature that relates the atmospheric temperature profile at a given time
to the dependence of muon production on atmospheric depth.  This paper aims to review and generalize the treatment of muon production and effective temperature
that has been used to interpret seasonal variations of atmospheric muons by many
experiments.  The formalism is developed both in integral form -- for
application to compact detectors at a fixed depth that record all muons with
$\Emu > \Emumin$ -- and in differential form -- for application to extended detectors 
like IceCube, KM3NeT, and Baikal-GVD, where the rates are proportional to energy-dependent effective areas.
\end{abstract}




\end{frontmatter}


\noindent \textit{This paper combines and builds upon the work on seasonal variations of atmospheric muons performed by Thomas~K.~Gaisser together with various collaborators over the last decade. It is partly based on and expanded from two paper drafts which were in preparation but which remain unfinished due to his passing.}

\section{Introduction}\label{sec:intro}

Atmospheric muons come primarily from the decay of charged pions and kaons produced by
cosmic-ray interactions in the upper atmosphere.  In the energy range where the 
interaction lengths of the parent mesons are comparable to their decay lengths,
higher temperatures lead to lower density and, therefore, to higher muon production rates.  The degree of correlation evolves over an energy
range defined by the critical energies for pions $\epsilon_\pi \approx 115$~GeV 
and kaons $\epsilon_K \approx 857$~GeV, where the numerical values correspond to a temperature of \SI{220}{\kelvin}.  The correlation with temperature 
is small at low energies, $E_\mu < \epsilon_\pi$, where most mesons decay, and becomes fully correlated
for muons energies above several TeV.  Because of the difference in their critical energies, the
$\pi^\pm/K^\pm$ production ratio is an important factor in this study.  
Prompt muons from the decay of charmed hadrons and neutral vector mesons remain uncorrelated
with temperature below their critical energies $\sim 10^7$~GeV~\cite{Desiati:2010wt}, but make a
negligible contribution to the overall rates and are therefore not considered. 

Seasonal variation of atmospheric muons has been a benchmark measurement of every underground
detector since the classic paper~\cite{Barrett:1952woo} on muons in a salt mine near Ithaca,
New York. At a depth of 1574 m.w.e., muons in that detector required $\Emu > \SI{440}{\giga\eV}$ at production
to reach the detector. Measurements with experiments at
the Laboratori Nazionali del Gran Sasso (LNGS), starting with
MACRO~\cite{Ambrosio:1997tc} and LVD~\cite{Vigorito:2019evi}, 
have a variable overburden corresponding approximately to $\Emu > 1.5\pm \SI{0.2}{\tera\eV}$ at production
depending on the exact location of the detector.  More
recent observations at LNGS include BOREXINO~\cite{Bellini:2012te}, GERDA~\cite{Agostini:2016gda}, and 
OPERA~\cite{Agafonova:2018kce}.  The MINOS far detector at a depth of 2100 m.w.e. 
in the Soudan mine~\cite{Adamson:2009zf} detects muons with $\Emu > \SI{730}{\giga\eV}$ at production.  There
are also measurements with shallower experiments such as the MINOS near 
detector~\cite{Adamson:2014xga,Adamson:2015qua} and
NOvA~\cite{Acero:2019lmp} at Fermilab that correspond to $\Emu \gtrsim \SI{50}{\giga\eV}$.

The relation between measured muon rate $R$ and atmospheric temperature is conventionally
quantified by a correlation coefficient, $\alpha_T$,
\begin{equation}
\frac{\Delta R}{\langle R \rangle} = \alpha_T\frac{\Delta T_\mathrm{eff}}{\langle T_\mathrm{eff} \rangle},
\label{eq:alpha_T}
\end{equation}
where $\Teff$ is the effective temperature characterizing the atmospheric temperature profile. The $\Delta$ in \refeq{eq:alpha_T} indicates the variation with respect to the yearly average muon rate $\langle R \rangle$ and effective temperature $\langle T_\mathrm{eff} \rangle$. Several experimental measurements of the temperature correlation coefficient show that it varies from $0.2$ to $0.95$ in the energy range from \SI{20}{\giga\eV} to $\sim\mathcal{O}(\rm{TeV})$~\cite{Agafonova:2018kce}.

This paper is organized with an initial section relating the
muon rate at the detector to the production spectrum of muons as a function of the atmospheric depth, both for
compact detectors like MINOS and those at LNGS,
and for the deep neutrino telescopes that span a large range of depths. The focus is on an analytic approximation for the muon production spectrum, but two alternative approaches are considered. The next 
section relates the muon production spectrum to weights for calculating the effective
temperature by weighting the temperature profile at each depth.  It also
includes a comparison of 
the weights of this work with those defined by Grashorn et al.\ in Ref.~\cite{Grashorn:2009ey} 
and used by many measurements. This is followed by a discussion on the correlation
coefficient and its dependence on energy and zenith angle. The following section demonstrates how the correlation coefficient is a probe of the atmospheric $K/\pi$-ratio. Finally, we comment on the impact of multiple muon events and nuclear primaries on the rate calculation.
An Appendix provides details of the hadronic cascade
equations and their approximate solutions with details of how the lepton spectra
are calculated.
\section{Muon rate  calculations}\label{sec:rate}

The evolution of a cascade of particles in the atmosphere can be described by the coupled cascade equations~\cite{Gaisser:2016uoy} 
\begin{equation}
\begin{split}
\frac{\mathrm{d} \phi_i(E,X)}{\mathrm{d} X}  =
 &-\frac{\phi_i(E,X)}{\lambda_{\text{int}, i}(E)} -\frac{\phi_i(E,X)}{\lambda_{\text{dec},i}(E, X)}\\
 &+ \sum_j{\int_E^\infty \mathrm{d} E_j~ \frac{\mathrm{d} n_{j(E_j) \to i(E)}}{\mathrm{d} E} 
                          \frac{\phi_j(E_j,X)}{\lambda_{\text{int}, j}(E_j)}}\\
 &+ \sum_j{\int_E^\infty \mathrm{d} E_j~ \frac{\mathrm{d} n^\text{\text{dec}}_{j(E_j) \to i(E)}}{\mathrm{d} E} 
                          \frac{\phi_j(E_j,X)}{\lambda_{\text{dec}, j}(E_j, X)}}.
\end{split}
\label{eq:cascade_equations}
\end{equation}
Here, $\phi_i(E, X) \,\dd E$ is the flux of particles of type $i$ at atmospheric slant depth $X$ with energies in the interval $E$ to $E + \dd E$. The first two terms on the right-hand side of \refeq{eq:cascade_equations} are loss terms as a result of interaction and decay of particles $i$, governed by the interaction and decay lengths $\lambda_\mathrm{int}$ and $\lambda_\mathrm{dec}$. The last two terms are source terms for the production of particle type $i$ due to the interaction and decay of particles of type $j$, where $\mathrm{d}n/\mathrm{d}E$ are the inclusive particle production spectra. For an observation height $h_0$ in the atmosphere, the slant depth $X$ in Eq.~(\ref{eq:cascade_equations}) is given along the trajectory $l$ of the central core of the cascade by
\begin{equation}
    X(h_0, \theta)=\int_{h_0}^\infty dl\; \rho_\mathrm{air}(h(l,\theta)),
    \label{eq:density}
\end{equation}
where the mass density of air, $\rho_\mathrm{air}$, is typically a function of the atmospheric height $h(l,\theta)$, and $\theta$ is the zenith angle of the trajectory. Because the density is directly related to temperature, the fluxes of particles in air showers are sensitive to temperature fluctuations in the atmosphere.

The inclusive production spectrum of muons, differential in muon energy $E_\mu$ and atmospheric slant depth $X$, is then given by
\begin{equation}
    P(\Emu,\theta,X)=\frac{\dd \phi_\mu(\Emu,\theta,X)}{\dd X},
\end{equation}
when solving the cascade equations starting from the total primary nucleon flux $\phi_N$ at the top of the atmosphere. The flux of muons differential in energy at the surface is obtained from the integral over the production spectrum,
\begin{equation}
    \phi_\mu(E_\mu, \theta) = \int_0^{X_O} \dd X P(\Emu, \theta, X).
\end{equation}
Due to the relation to the atmospheric density profile, the muon production spectrum implicitly depends on the temperature $T(X)$ at slant depth $X$. 

The rate of muons with energy $\Emu$ from a direction corresponding to a zenith angle $\theta$ in a detector with effective area $A_\mathrm{eff}(\Emu,\theta)$ is given by
\begin{align}
R(\theta) &= \int \dd X\int_{\Emumin}^\infty \dd\Emu\, \Aeff(\Emu,\theta)\,P(\Emu,\theta,X).
\label{eq:rtheta}
\end{align}
For a compact detector at a depth large compared to its vertical dimension, the effective area is simply its projected physical area at the zenith angle $\theta$ averaged over the azimuth angle. In this case,
\begin{equation}
\label{eq:compact}
\begin{split}
R(\theta)& =\Aeff(\theta) \int{\rm d}X\int_{\Emumin}{\rm d}\Emu\,P(\Emu,\theta,X) \\
& = \Aeff(\theta)\int{\rm d}X\,\Pint(\Emumin, \theta, X) \\
& = \Aeff(\theta)\,I(\Emumin,\theta),
\end{split}
\end{equation}
where $I(\Emumin,\theta)$ is the integral muon flux for $\theta$, $\Pint(\Emumin, \theta, X)$ is the integral version of the production profile\footnote{In previous works, the integral muon production spectrum has sometimes been written as $P(>E_\mu, \theta, X)$, which represents the differential production spectrum $P$ integrated over all energies above some minimum energy $E_\mu$. For clarity, we choose in this work to use instead the notation $\Pint(\Emumin, \theta, X) \equiv \int_{\Emumin}^{+\infty} P(\Emu, \theta, X) \dd \Emu$.}, and \Emumin is the energy threshold for a muon to reach the detector at this angle.
In both cases, the total rate, $R$, is given by integrating over the solid angle $\Omega$,
\begin{equation}
R = \int R(\theta)\,\dd\Omega.
\label{eq:rate}
\end{equation}
The differential version (Eq.~\ref{eq:rtheta}) is
appropriate for a geometrically extended experiment like IceCube where the effective area depends on
muon energy, for example, because higher energy is required for a muon at a large angle to reach the lower part of the detector. Furthermore, such experiments are sparsely instrumented which may cause a fraction of muons to fail to pass the trigger or subsequent analysis steps, an effect which usually diminishes with increasing energy.  For a compact detector at a given depth,  e.g. MACRO, MINOS, and NOvA,
any muon with sufficient energy to reach the depth of the detector can be recorded if it passes
through the detector.  In this case, the integral version of the production profile
as in Eq.~\ref{eq:compact} is
appropriate (and has been used traditionally).

In the following sections, three approaches to obtaining the muon production spectrum are described. The first approach consists of an approximate analytical solution to the cascade equations including the pion and kaon channels. A second approach utilizes a numerical solver of the cascade equations which includes all relevant channels. A third and conceptually different approach is based on a parameterization of muon production profiles in extensive air showers, which are integrated over the flux of primary particles. For the purpose of illustration, a hypothetical cylindrical detector with a radius of \SI{5}{\m} and a height of \SI{20}{\m} at a depth of \SI{2000}{\mwe} is used. For a compact detector at such a large depth, the effective area is given by the projected physical area. The average minimum energy that a muon requires to reach the detector is estimated from the approximation given in \refref{Gaisser:2016uoy}. We consider these values as a sharp cutoff above which muons are detected and below which they are not\footnote{Note that because of the steeply falling spectrum of primary nucleons and consequently of atmospheric muons, an accurate description of the threshold region is crucial for accurate rate calculations for real detectors.}. The numerical values of the effective areas and threshold energies used in the calculations are given in \reftab{tab:detector}. The muon rates are calculated using daily temperature data of the South Pole atmosphere for the year 2012 obtained from the Atmospheric Infrared Sounder (AIRS) on board of NASA's Aqua satellite~\cite{AIRS2013}. AIRS is capable of measuring the geopotential height and temperature in the atmosphere with an accuracy of \SI{1}{\kelvin} over 24 pressure layers between \SI{1}{hPa} and \SI{1000}{hPa}, even under cloudy conditions. Assuming an ideal gas law, the corresponding atmospheric density profile, $\rho_\mathrm{air}$, can be obtained using the AIRS pressure and temperature data. All calculations are performed using the primary nucleon spectrum from Tom Gaisser's H3a flux~\cite{Gaisser:2011klf}.

\begin{table}[tb]
\centering
\setlength{\tabcolsep}{4pt}
\footnotesize
\begin{tabular}{l|llllllll}
$\cos(\theta)$            & 0.95 & 0.85 & 0.75 & 0.65 & 0.55 & 0.45 & 0.35 & 0.25  \\
\hline
\Emumin (GeV)   & 660  & 781  & 952  & 1211 & 1641 & 2458 & 4416 & 11766 \\
$A_\mathrm{eff}$ (m$^2$) & 137  & 172  & 191  & 203  & 210  & 214  & 215  & 213   \\
\end{tabular}
\caption{Effective area and threshold energy for muons at the Earth's surface to reach a hypothetical cylindrical detector of radius \SI{5}{\m}, height \SI{20}{\m}, and depth \SI{2000}{\mwe}.}
\label{tab:detector}
\end{table}

\subsection{Approximate analytical solution of the cascade equations}\label{sec:analytic}

The differential production profiles obtained from the cascade equations in the limits of low and high energy 
are repeated here from Ref.~\cite{Gaisser:2016uoy} and presented in detail
in \ref{sec:appendix} of this paper. The low- and high-energy regime is defined relative to the critical energies of the parent mesons of the muons, given by
\begin{equation}\label{eq:epsilon_pi}
    \epsilon_\pi = \frac{m_\pi c^2}{c \tau_\pi} \frac{RT}{Mg} \approx \SI{115}{\giga\eV}\, \frac{T}{\SI{220}{\kelvin}}
\end{equation}
for pions, and equivalent for $\epsilon_K$. Here, $m_\pi$ and $\tau_\pi$ are the mass and lifetime of the pion, $g$ is the acceleration of free fall, $R$ the molar gas constant, $M$ the mean molar mass for air, and $T$ the atmospheric temperature. For muons with $\Emu \lll \epsilon_\pi$,
\begin{multline}
\label{eq:mu-prod-low}
P(\Emu,\theta,X)\approx \phi_N(\Emu)\,\frac{e^{-X/\Lambda_N}}{\lambda_N}\\
\times\left[ \frac{Z_{N\pi}(1-r_\pi^{\gamma+1})}{(\gamma+1)(1-r_\pi)} +
0.636\,\frac{Z_{NK}(1-r_K^{\gamma+1})}{(\gamma+1)(1-r_K)}\right ],
\end{multline}
and for muons with $\Emu \ggg \epsilon_K$
\begin{align}
\label{eq:mu-prod-high}
P(\Emu,\theta,X)&\approx \phi_N(\Emu) \times \Big[ \nonumber \\ &\frac{\epsilon_\pi}{X\cos(\theta)\Emu}
\frac{(1-r_\pi^{\gamma+2})}{(1-r_\pi)(\gamma+2)}
\frac{Z_{N\pi}}{1-Z_{NN}}\frac{\Lambda_\pi}{\Lambda_\pi - \Lambda_N}
\nonumber \\ \times
&\left(e^{-X/\Lambda_\pi}-e^{-X/\Lambda_N}\right)
\nonumber \\ + 0.636 &\frac{\epsilon_K}{X\cos(\theta)\Emu}
\frac{(1-r_K^{\gamma+2})}{(1-r_K)(\gamma+2)} 
\frac{Z_{NK}}{1-Z_{NN}}\frac{\Lambda_K}{\Lambda_K - \Lambda_N}
\nonumber \\ \times
&\left(e^{-X/\Lambda_K}-e^{-X/\Lambda_N}\right)\Big],
\end{align}
where $r_\pi = m_\mu^2/m_\pi^2$, and $\lambda$ and $\Lambda$ are atmospheric interaction and attenuation lengths respectively. These equations are obtained by integrating
solutions of the hadronic cascade equations (\refeq{eq:cascade_equations}) for charged pions and kaons to get the
spectrum of leptons from $\pi^\pm / K^\pm \rightarrow \mu^\pm + \nu_\mu (\bar\nu_\mu)$, given a primary nucleon flux \mbox{$\phi_N(E) \propto E^{-(\gamma + 1)}$}, with $\gamma$ the integral spectral index.
The integral over the primary flux is related to the primary flux evaluated at the energy of the muon by spectrum-weighted moments $Z_{Nh}$. The $Z$-factors are given by
\begin{equation}\label{eq:Z}
    Z_{Nh} = \int_0^1 x^\gamma \frac{\dd n_{N\rightarrow h}}{\dd x} \dd x,
\end{equation}
where $x=E_h / E_N$. This definition assumes Feynman scaling for the particle production and a constant spectral index $\gamma$, so that the spectrum-weighted moments are constants. Such an approximation is realistic
because of the steepness of the primary spectrum and the threshold of the deep
detector, which combine to limit the range of relevant primary energies.

An approximation valid for all energies can be obtained with the form
\begin{equation}
P(\Emu,\theta,X)=\frac{\rm Low}{1 + {\rm Low}/{\rm High}},
\label{eq:low-high}
\end{equation}
where \emph{Low} refers to \refeq{eq:mu-prod-low} and \emph{High} refers to \refeq{eq:mu-prod-high}. The approximations are made separately for pions and kaons.
From \refeq{eq:mu-prod-low} we see that
\begin{equation}
P_{\pi}(X) = \frac{A_{\pi\mu}(X)}{1+B_{\pi\mu}(X)\Emu\cos(\theta)/\epsilon_\pi(T)}
\label{eq:piofX}
\end{equation}
with
\begin{equation}
A_{\pi\mu}(X)=\frac{Z_{N\pi}}{\lambda_N(\gamma+1)}\frac{1-r_\pi^{\gamma+1}}{1-r_\pi}e^{-X/\Lambda_N},
\label{eq:Api}
\end{equation}
and from \refeq{eq:mu-prod-high}
\begin{equation}
B_{\pi\mu}(X)=\frac{\gamma+2}{\gamma+1}\,\frac{1-r_\pi^{\gamma+1}}{1-r_\pi^{\gamma+2}}\,\frac{X}{\Lambda^*}
\,\frac{e^{-X/\Lambda_N}}{e^{-X/\Lambda_\pi}-e^{-X/\Lambda_N}},
\label{eq:Bpi}
\end{equation}
where $\Lambda_\pi^* = \Lambda_\pi\times\Lambda_N/(\Lambda_\pi-\Lambda_N)$ is a combination of the 
attenuation lengths for nucleons and pions.
The equations for the kaon channel have the same form with $A_{K\mu}(X)$ multiplied by a factor of 0.636, the branching ratio for the decay $K^\pm \rightarrow \mu^\pm + \nu_\mu(\bar\nu_\mu)$~\cite{ParticleDataGroup:2020ssz}.
The total differential production spectrum is
\begin{equation}
P(\Emu,\theta,X)=\phi_N(\Emu)\left\{P_{\pi}(X)+P_{K}(X)\right\},
\label{eq:Panalytic}
\end{equation}

\begin{figure}[tb]
    \centering
    \includegraphics[width=\linewidth]{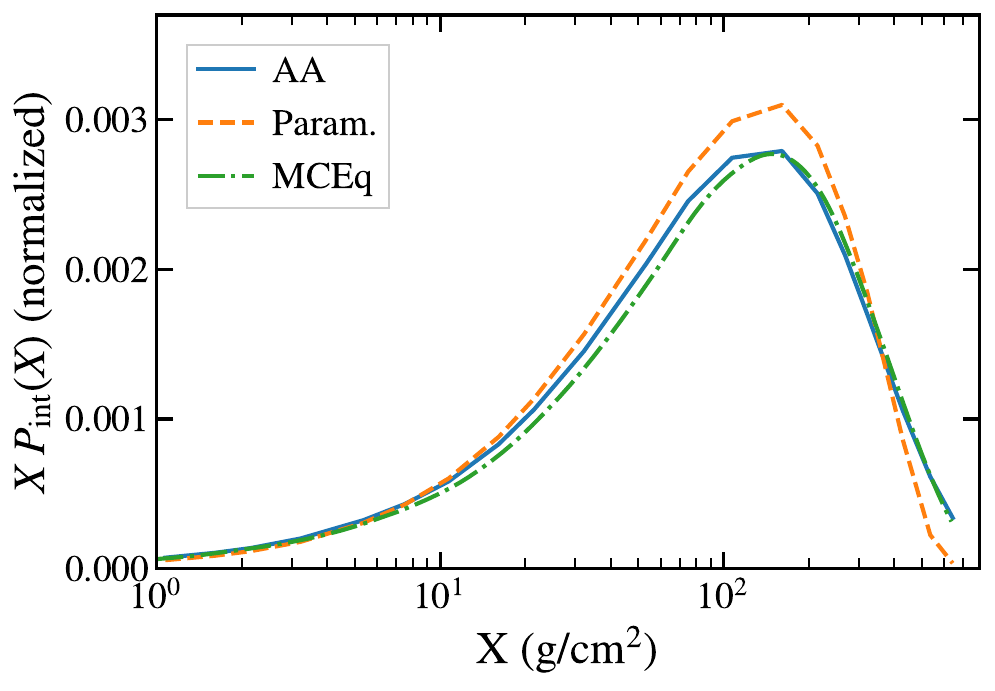}
    \caption{Integral muon production profiles obtained for $\Emu > \SI{500}{\giga\eV}$ and $\cos(\theta)=0.95$ using three different methods: the analytic approximation (AA, \refsec{sec:analytic}), the numerical solver (MCEq, \refsec{sec:mceq}), and the parameterized air-shower production profiles integrated over the primary spectrum (Param., \refsec{sec:param}).}
    \label{fig:Pmu}
\end{figure}

The equations assume a power law primary spectrum, where 
$\phi_N(\Emu) = C_N\times \Emu^{-(\gamma+1)}$ is primary nucleons per
\mbox{GeV m$^2$s sr}.  When the low-energy 
form \refeq{eq:mu-prod-low} is integrated to get the corresponding integral production profile, 
\mbox{$\phi_N(\Emu)\rightarrow E\times \phi_N(\Emu)/\gamma$}.  The high-energy form (\refeq{eq:mu-prod-high}) has an additional 
factor of muon energy in the denominator, so \mbox{$\phi_N(\Emu)\rightarrow E\times \phi_N(\Emu)/(\gamma+1)$}
at high energy.  Applying the approximation of Eq.~(\ref{eq:low-high}) then leads to
\begin{multline}
\Pint(\Emumin,\theta,X) = \Emumin \phi_N(\Emumin)\\\times \frac{A_{\pi\mu}(X)}{\gamma+(\gamma+1)B_{\pi\mu}(X)\Emumin\cos(\theta)/\epsilon_\pi}.
\label{eq:integralPmu}
\end{multline}
This form (plus the corresponding term for kaons) 
provides the production profile that can be inserted into Eq.~\ref{eq:compact} to get the inclusive rate of muons (assuming an \Aeff that does not depend on muon energy). The production profile for a specific \Emu and $\cos(\theta)$ is shown in \reffig{fig:Pmu}.

The above equations are for $\mu^+ +\mu^-$.  The corresponding equations for $\nu_\mu+\overline{\nu}_\mu$
have the same form with the meson decay kinematic factors like $(1-r_\pi^{\gamma+1})$ and
$(1-r_K^{\gamma+2})$ replaced by $(1-r_\pi)^{\gamma + 1}$ and $(1-r_K)^{\gamma+2}$, respectively~\cite{Gaisser:2019abc}.

The constants used in the calculations are given in \reftab{tab:constants}. More detail can be included in the calculation by taking into account
the non-scaling behavior of hadronic interactions and gradual changes of the primary
spectral index. To first approximation, this is done by introducing energy-dependent spectrum-weighted moments as in \refref{Gondolo:1995fq}. For this work, we compared a calculation using the constant values from \refref{Gaisser:2016uoy} based on Sibyll~2.3 (\reftab{tab:constants}), and a calculation using energy-dependent values obtained from Sibyll~2.3c~\cite{Riehn:2019jet} (see \reffig{fig:not-constants}). While the calculation with energy-dependent values gives a higher rate, the difference is nearly constant with the relative variations throughout the year deviating by less than 0.1\% (see \ref{sec:appendix}). In \reffig{fig:rates}, we show the daily rate calculated with the energy dependent parameters, compared to the rates obtained with the other methods considered. The calculated angular distribution of the events is shown in \reffig{fig:angular}.

It is possible to check the accuracy of \refeqs{eq:low-high}{eq:piofX} by 
expanding the exact solution of the cascade equations in \refeq{eq:exact}. A comparison of predictions given by the analytical approximation described here to a full numerical solution as described in the following section was presented earlier in \refref{Gaisser:2019xlw}.

\subsection{Numerical solution of cascade equations}\label{sec:mceq}

The approximate analytical solutions of the cascade equations are based on various simplifications that can introduce uncertainties on the atmospheric muon fluxes. In order to estimate these uncertainties Monte-Carlo simulations or numerical solutions of the
coupled cascade equations are required. The software package MCEq (Matrix Cascade Equations)~\cite{Fedynitch:2015zma} provides precise numerical solutions of the cascade equations with a level of detail comparable with current Monte Carlo simulations. To achieve this, the cascade equations are expressed in matrix form to make use of modern implementations of linear algebra algorithms. The calculations rely on several input parameters, such as the initial cosmic-ray flux and the atmospheric density profile. Further details can be found in Ref.~\cite{Fedynitch:2015zma}. An extension of this approach is realized with the Muon Intensity Code (MUTE)~\cite{Fedynitch:2021ima} which accounts for muon propagation in dense media to estimate muon fluxes in deep-underground experiments. However, in this work the simple approach based on effective areas and energy thresholds, as described in Section~\ref{sec:analytic} (Table~\ref{tab:detector}), is used to obtain the expected muon flux in a hypothetical cylindrical detector of radius \SI{5}{\m}, height \SI{20}{\m}, and depth \SI{2000}{\mwe}.

\begin{figure}[tb]
    \centering
    \includegraphics[width=\linewidth]{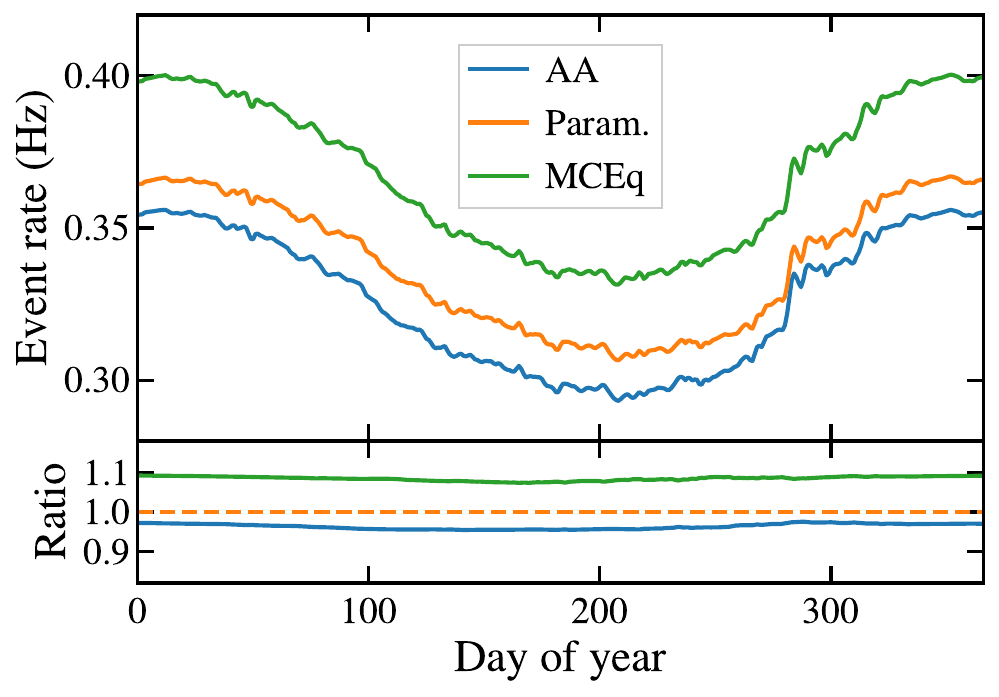}
    \caption{Daily event rate for the detector of \reftab{tab:detector} calculated following the three different methods from \refsecss{sec:analytic}{sec:mceq}{sec:param}. (The sharp increase in the expected rate around October is a feature of the South Pole atmosphere, see for example \refref{Tilav:2019xmf}.)}
    \label{fig:rates}
\end{figure}

\begin{figure}[tb]
    \centering
    \includegraphics[width=\linewidth]{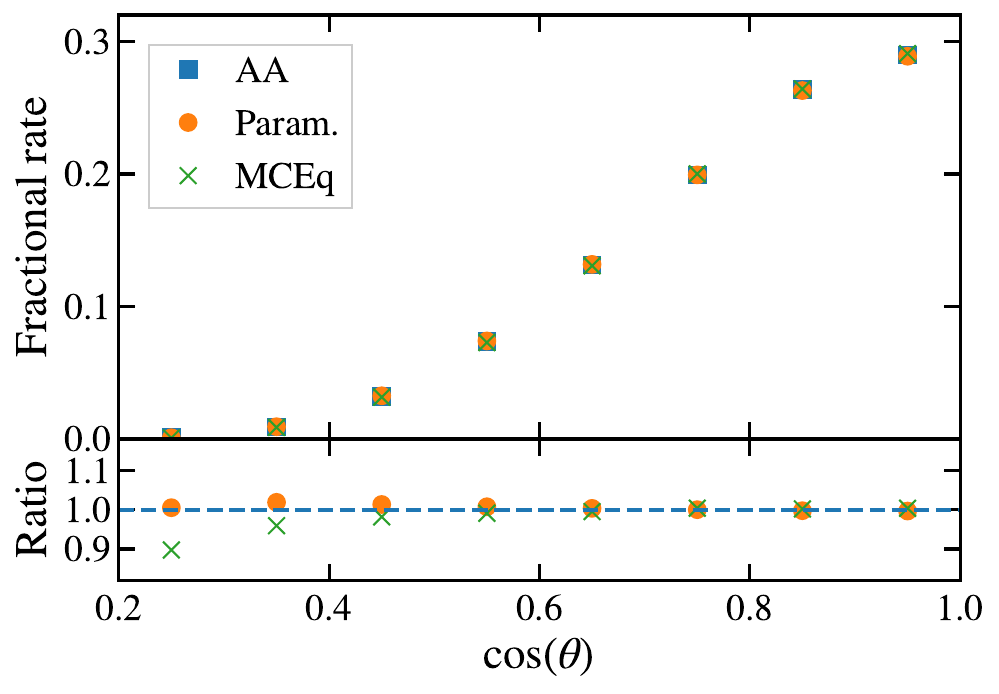}
    \caption{Expected zenith angle distribution of muons calculated for the detector of \reftab{tab:detector} following the three different methods from \refsecss{sec:analytic}{sec:mceq}{sec:param}. (For the AA and Param. calculations, the effect of the curvature of the Earth was approximated by replacing $\cos(\theta)$ in the formula by an effective $\cos(\theta^*)$ from \refref{Chirkin:2004ic}.)}
    \label{fig:angular}
\end{figure}

The atmospheric muon flux is determined with MCEq, using Sibyll 2.3c, at different atmospheric depths, $X_i$, assuming the primary nucleon flux from H3a and a daily atmospheric temperature and density profile at the South Pole derived from 2012 AIRS data. Subtracting the muon spectrum at $X_{i+1}$ from the spectrum at $X_i$ for all $i$ then directly yields the muon production spectrum $P(\Emu,\theta,X)$, which is shown in \reffig{fig:Pmu}. The expected muon rate, $R(\theta)$, is then calculated according to Eq.~(\ref{eq:rtheta}). Analogously to the analytical approach, integration over the solid angle yields the total muon rate in the detector, as described in Eqs.~(\ref{eq:compact}) and (\ref{eq:rate}). The resulting total muon rate is shown in \reffig{fig:rates} and the corresponding angular distribution in \reffig{fig:angular}.

\subsection{Parameterization of Monte Carlo cascades}\label{sec:param}

An alternative approach consists of integrating average muon production profiles in air showers over the primary cosmic-ray flux. A parameterization of such profiles based on simulations and its applications are described in \refref{Gaisser:2021cqh}. The differential muon production spectrum
$P(\Emu,\theta,X)$ is given by 
\begin{equation}
P(\Emu,\theta,X)=\int_{\Emu}g(\Emu,E_0,\theta,X, T)\,\phi_N(E_0){\rm d}E_0
\label{eq:PofEmu}
\end{equation}
with
\begin{equation}
g(\Emu,E_0,\theta,X, T) = \frac{\rm d}{{\rm d}\Emu}\left(\frac{{\rm d}N(\Emumin,E_0,A,\theta,X,T)}{{\rm d}X}\right).
\label{eq:params}
\end{equation}

Here $\mathrm{d}N(\Emumin,E_0,A,\theta,X,T)/{\rm d}X$ 
is the mean number of muons with energy $>\Emumin$ produced per g/cm$^2$
as a function of slant depth $X$ in a cosmic-ray air shower initiated by a primary nucleus with mass number $A$, primary energy $E_0$, and zenith angle $\theta$, where the atmospheric temperature at $X$ is given by $T$.  It is a product of
the derivative of the Gaisser-Hillas (\mbox{G-H}) function~\cite{GaisserHillasabc}, often used to fit the longitudinal development of air showers and its derivative here interpreted as the longitudinal production of mesons in the cascade,
multiplied by a decay factor that provides the temperature
dependence of the decay probability of pions and kaons to muons, and a threshold factor:
\begin{equation}
\begin{split}
\label{eq:formula} 
\frac{\dd N}{\dd X}(\Emumin,E_0,A&,\theta,X,T)= \\
&N_{\rm max}\,\exp((X_{\rm max}-X)/\lambda)\\ 
\times\,&\frac{X_{\rm max}-X}{\lambda\,(X-X_0)} \,\left(\frac{X-X_0}{X_{\rm max}-X_0}\right)^{(X_{\rm max}-X_0)/\lambda} \\
\times\,& F(\Emumin,T) \,\left(1-\frac{A\Emumin}{E_0}\right)^{5.99}.
\end{split}
\end{equation}
The parameters $N_\mathrm{max}$, $X_\mathrm{max}$, $\lambda$, $X_0$ are the free parameters appearing in the original \mbox{G-H} function, which were parameterized in \refref{Gaisser:2021cqh} in terms of $E_0$, $A$, and \Emumin based on fits to muon production profiles obtained from air-shower simulations. For the parameterization,
a scaling form depending on $E_0/(A\,\Emumin)$ is used so that only the primary
spectrum of nucleons is required in Eqs.~\ref{eq:params} etc.

The decay factor is 
\begin{equation}
F(\Emumin,T) = \frac{f_\pi}{1+\frac{(f\Emumin)\cos(\theta) X}{r_\pi\lambda_\pi \epsilon_\pi(T)}}
+\frac{f_K}{1+\frac{(f\Emumin)\cos(\theta) X}{r_K\lambda_K \epsilon_K(T)}},
\label{eq:Fdecay}
\end{equation}
with $f\ge 1$, a factor fitted from the simulations that gives the mean energy of all muons with energy greater than $\Emumin$;
$r_\pi=0.79$ and \mbox{$r_K=0.52$} are the fraction of the parent meson momentum carried by
the muon, and $\lambda_\pi=\SI{110}{\g\per\cm\squared}$ and $\lambda_K=\SI{122}{\g\per\cm\squared}$ are the meson interaction lengths.

The normalization factors $f_\pi$ and $f_K$ of the pion and kaon component are defined in terms of the average momentum they carry away in interactions of nucleons in the atmosphere, taking into account the branching ratio for the muon decay channel for charged kaons. This average momentum fraction is equivalent to the spectrum-weighted moment of \refeq{eq:Z} evaluated for \mbox{$\gamma=1$}. Requiring the sum of the normalization factors to be equal to one, they are defined as \mbox{$f_\pi=Z_{N\pi}^{\gamma = 1} / (Z_{N\pi}^{\gamma = 1} + 0.636\times Z_{NK}^{\gamma=1}) = 0.92$} and \mbox{$f_K = 1 - f_\pi = 0.08$}, where numerical values from \refref{Gaisser:2016uoy} were used for $Z^{\gamma=1}$.

The inclusive muon production profile calculated according to \refeqs{eq:PofEmu}{eq:params} is shown in \reffig{fig:Pmu}. The calculated rates and zenith distribution are shown in \reffigs{fig:rates}{fig:angular}. For our calculations, we use the fit parameters given in Table~1 from \refref{Gaisser:2021cqh} for the four functions fitted to Monte Carlo for $N_{\rm max}$, $X_{\rm max}$, $\lambda$, and $X_0$.

The integral over slant depth of Eq.~\ref{eq:formula} is equivalent to the Elbert formula~\cite{Elbert:1979gz,Elbert:1979abc}
approximation for the average number of muons per shower for a given zenith angle~\cite{Gaisser:1985yw}:
\begin{equation}
\langle N(\Emumin)\rangle\;\approx\;A\,\frac{K}{ \Emumin\cos(\theta)}\,\left (
\frac{E_0}{ A\Emumin}\right )^{\alpha_1}\,
\left (1-\frac{A\Emumin}{E_0}\right)^{\alpha_2},
\label{eq:ElbertFormula}
\end{equation}
where $A$ is the mass number of a primary nucleus of total energy $E_0$\footnote{The Elbert formula traditionally uses the notation $\langle N_\mu(>\Emu)\rangle$, written here instead as $\langle N(\Emumin)\rangle$.}.
The dependence on the ratio $A\,\Emumin/E_0$ follows from the superposition approximation, in which
incident nuclei are treated as $A$ independent nucleons each of energy $E_0/A$. The threshold factor, i.e. the last factor in \refeq{eq:ElbertFormula}, is the same as for Eq.~\ref{eq:formula}. The benefit of integrating \refeq{eq:formula} over \refeq{eq:ElbertFormula} is the dependence on atmospheric temperature of the former.

Comparisons between the approach presented in this section and the analytic calculation from \refsec{sec:analytic} were shown earlier in \refref{Gaisser:2021bwj}. Alternatively to the parameterization of production profiles obtained from simulation as discussed in this section, one could use MCEq (\refsec{sec:mceq}) to obtain average production profiles in air showers by using it to solve the cascade equations with a single primary particle as the initial condition. This will increase the accuracy of the calculation in various ways, for example, by taking into account all relevant muon production channels and including the energy dependence of the inclusive cross sections, as well as through the implementation of the curved geometry relevant for more horizontal directions.

\section{Effective temperature}\label{sec:Teff}

The variation of muon rate with atmospheric conditions is ordinarily described in terms of correlation with an effective temperature parameter. The effective temperature characterizes the atmospheric temperature profile by averaging it with appropriate weights related to the muon production spectrum. Different definitions of effective temperature have been used in the literature.

\begin{figure}[tb]
    \centering
    \includegraphics[width=\linewidth]{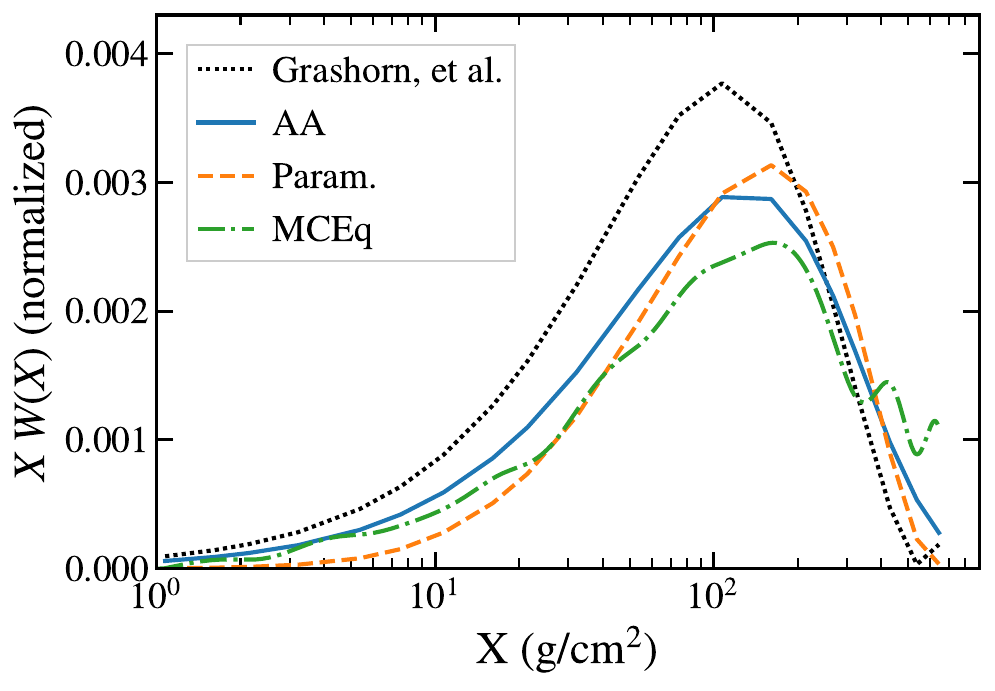}
    \caption{Weight of different atmospheric depths in the calculation of the effective temperature according to \refeq{eq:Teff}. The black dotted line show the (temperature-independent) values from \refref{Grashorn:2009ey}. They are compared to values obtained using the methods of \refsec{sec:rate}, averaged over a full year, as these weights depend on the atmospheric temperature profile. The calculations are done for \Emumin = \SI{500}{\giga\eV} and $\cos(\theta)=0.95$.}
    \label{fig:weights}
\end{figure}

A definition that has traditionally been used can be obtained by taking the derivative of the rate in \refeq{eq:rate} with respect to temperature. The change in rate is obtained by integrating the change in atmospheric temperature over depth, i.e.
\begin{equation}
    \Delta R(\theta) = \int \mathrm{d}X\int\mathrm{d}\Emu A_\mathrm{eff}(\Emu, \theta) \, \frac{\mathrm{d}P(\Emu, \theta, X)}{\mathrm{d}T} \Delta T(X).
\end{equation}
Defining $\Delta T(X) = T(X) - T_\mathrm{eff}$ and setting $\Delta R = 0$ for an isothermal atmosphere where $T(X) = T_\mathrm{eff}$ results in the following definition:
\begin{equation}
T_{\rm eff}(\theta) = \frac{\int{\rm d}X\int{\rm d}\Emu\, \Aeff(\Emu,\theta)
T(X)\frac{{\rm d}P(\Emu,\theta,X)}{{\rm d}T}}{\int{\rm d}X\int{\rm d}\Emu\, \Aeff(\Emu,\theta)
\frac{{\rm d}P(\Emu,\theta,X)}{{\rm d}T}}.
\label{eq:Teff}
\end{equation}
The total effective temperature is the weighted average of \refeq{eq:Teff} over the zenith distribution.
The corresponding integral form is
\begin{equation}\label{eq:Teff_integral}
    T_\mathrm{eff} (\theta) = \frac{\int \mathrm{d}X T(X) \frac{\mathrm{d}\Pint(\Emumin, \theta, X)}{\mathrm{d}T}}{\int \mathrm{d} X \frac{\mathrm{d}\Pint(\Emumin, \theta, X)}{\mathrm{d}T}}.
\end{equation}
It applies to compact detectors for which the effective area cancels in \refeq{eq:Teff} at each zenith angle.
For the analytic inclusive form of the pion channel spectrum from \refsec{sec:analytic}, for example,
\begin{equation}
    T(X) \frac{\mathrm{d} P (\Emu, \theta, X)}{\mathrm{d} T} = \frac{A_{\pi\mu}(X) B_{\pi\mu}(X) \Emu \cos(\theta) / \epsilon_\pi(T)}{\left[1 + B_{\pi\mu} \Emu \cos(\theta)/ \epsilon_\pi (T) \right]^2}.
\end{equation}
and
\begin{multline}\label{eq:TdPdT_int}
    T(X) \frac{\mathrm{d} \Pint(\Emumin, \theta, X)}{\mathrm{d} T} = \Emumin \phi_N(\Emumin) \\ \times\frac{A_{\pi\mu}(X) (\gamma +1) B_{\pi\mu}(X) \Emumin \cos(\theta)/\epsilon_\pi(T)}{\left[\gamma + (\gamma +1)B_{\pi\mu}(X)\Emumin\cos(\theta)/\epsilon_\pi(T)\right]^2}.
\end{multline}
An early implementation of this approach, presented in \refref{Grashorn:2009ey}, is used in the analysis of MINOS, among others. For comparison with the existing literature, it is necessary to write the effective temperature in terms of weights:
\begin{equation}\label{eq:Teffweights}
    T_\mathrm{eff}(\theta) = \frac{\int \dd X T(X) W(X)}{\int \mathrm{d}X W(X)} \approx \frac{\sum_i \delta \ln(X_i) T(X_i) X_i W(X_i)}{\sum_i \delta \ln(X_i) X_i W(X_i)}.
\end{equation}
The second form is motivated by the fact that atmospheric temperatures are commonly tabulated in quasi-logarithmic intervals of depth, so the integrations in this work are done logarithmically. From \refeq{eq:TdPdT_int}
\begin{multline}
    W(X) = \Emumin \phi_N(\Emumin) \\ \times \frac{A_{\pi\mu}(X) (\gamma +1) B_{\pi\mu}(X) \Emumin \cos(\theta)/\epsilon_\pi(T)}{T(X) \left[\gamma + (\gamma +1)B_{\pi\mu}(X)\Emumin\cos(\theta)/\epsilon_\pi(T)\right]^2}.
\end{multline}
The form obtained here differs from the one of \refref{Grashorn:2009ey}, with the weights now depending on the temperature profile through the critical energies. The normalized weights are compared in \reffig{fig:weights}. Despite the difference in the calculations, the weights are similar, with only a slight shift deeper in the atmosphere for the present calculation.

For the calculation of \Teff according to \refeq{eq:Teff}, with the parameterization of \refsec{sec:param}, the corresponding form for the decay factor is
\begin{equation}
    T(X) \frac{\mathrm{d}F(\Emumin, \theta, X)}{\mathrm{d}T} = \frac{f_\pi\, (f\Emumin) \cos(\theta) X / r_\pi \lambda_\pi \epsilon_\pi(T)}{\left[1 + (f\Emumin) \cos(\theta) X / r_\pi \lambda_\pi \epsilon_\pi (T)\right]^2}.
\end{equation}

To calculate the derivative of the muon production spectrum with respect to temperature with MCEq, first the production spectrum $P(\Emu,\theta,X)$ is determined as described in Sec.~\ref{sec:mceq}. In a second step, muon production spectra are derived for a local temperature change of $\dd T=\SI{1}{\kelvin}$. This is done by changing each atmospheric layer $i$ in the AIRS temperature and density profiles individually by \SI{1}{\kelvin} to obtain $\hat P_i(\Emu,\theta,X_i)$ and
\begin{equation}
    \hat P(\Emu,\theta,X)= \left(\hat P_1(\Emu,\theta,X_1), \dots, \hat P_n(\Emu,\theta,X_n)  \right),
\end{equation}
where $n$ is the total number of layers considered in the AIRS data.  The derivative of the production spectrum is then constructed via the difference quotient as
\begin{align}
    \frac{\dd P(\Emu,\theta,X)}{\dd T}&=\frac{P(T+\dd T)- P(T)}{\dd T} \nonumber\\ 
    &=\frac{\hat P(\Emu,\theta,X)-P(\Emu,\theta,X)}{\SI{1}{\kelvin}}.
\end{align}
The resulting derivative of the production spectrum in terms of weights, $W(X)$, is also shown in Fig.~\ref{fig:weights}.

Alternatively to the definition of \refeq{eq:Teff}, in which the atmospheric temperature profile is multiplied by the derivative of the muon production spectrum with respect to temperature, the effective temperature has been defined as a straightforward convolution of the temperature profile with the muon production spectrum, normalized to the muon rate for each angle~\cite{Desiati:2011hea}:
\begin{equation}
\tilde{T}_\mathrm{eff}(\theta) =\frac{\int{\rm d}X\,T(X)\int{\rm d}\Emu\, \Aeff(\Emu,\theta)P(\Emu,\theta,X)}{\int{\rm d}X\int{\rm d}\Emu\, \Aeff(\Emu,\theta)P(\Emu,\theta,X)}.
\label{eq:Tefftilde}
\end{equation}
\begin{figure}[tb]
    \vspace{-1em}

    \centering
    \includegraphics[width=\linewidth]{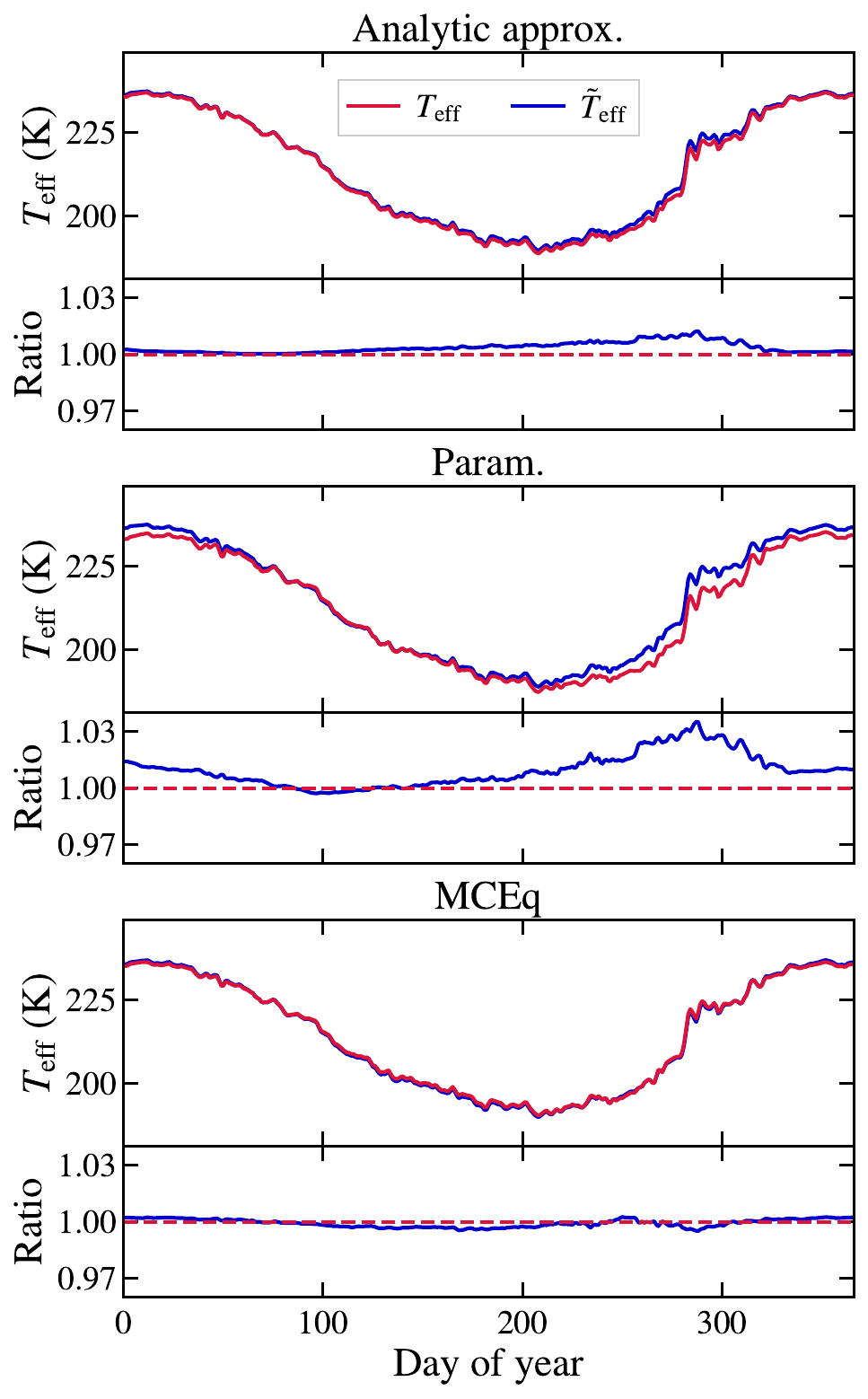}
    \caption{Comparison of the values of effective temperature obtained from the derivative definition \refeq{eq:Teff} and the alternative definition of \refeq{eq:Tefftilde} for the three different methods of calculation muon production discussed in \refsec{sec:rate}.}
    \label{fig:Teff_comparison}
    \vspace{-1em}
\end{figure}
A benefit of this definition is that the technical implementation is more simple compared to the derivative definition when using MCEq. A comparison of the daily effective temperature with the two definitions is shown in \reffig{fig:Teff_comparison}.

The relative variations in the calculated rate throughout the year are plotted as a function of relative variations of effective temperature in \reffig{fig:alpha_comparison}. The derivative definition of \Teff, \refeq{eq:Teff}, minimizes the difference between calculated rates on days that have the same value of \Teff. Using the alternative definition of \refeq{eq:Tefftilde}, a separation is visible between the months in which the atmosphere cools versus when it warms. This so-called hysteresis has been reported earlier by IceCube using this definition of effective temperature~\cite{Tilav:2019xmf}.
\section{Correlation coefficient}\label{sec:alpha}

The relation between the variation of effective temperature and the variation of muon rate can be expressed in terms of a correlation coefficient \aT as in \refeq{eq:alpha_T}.

\begin{figure}[tb]
    \centering
    \vspace{-1em}
    
    \includegraphics[width=\linewidth]{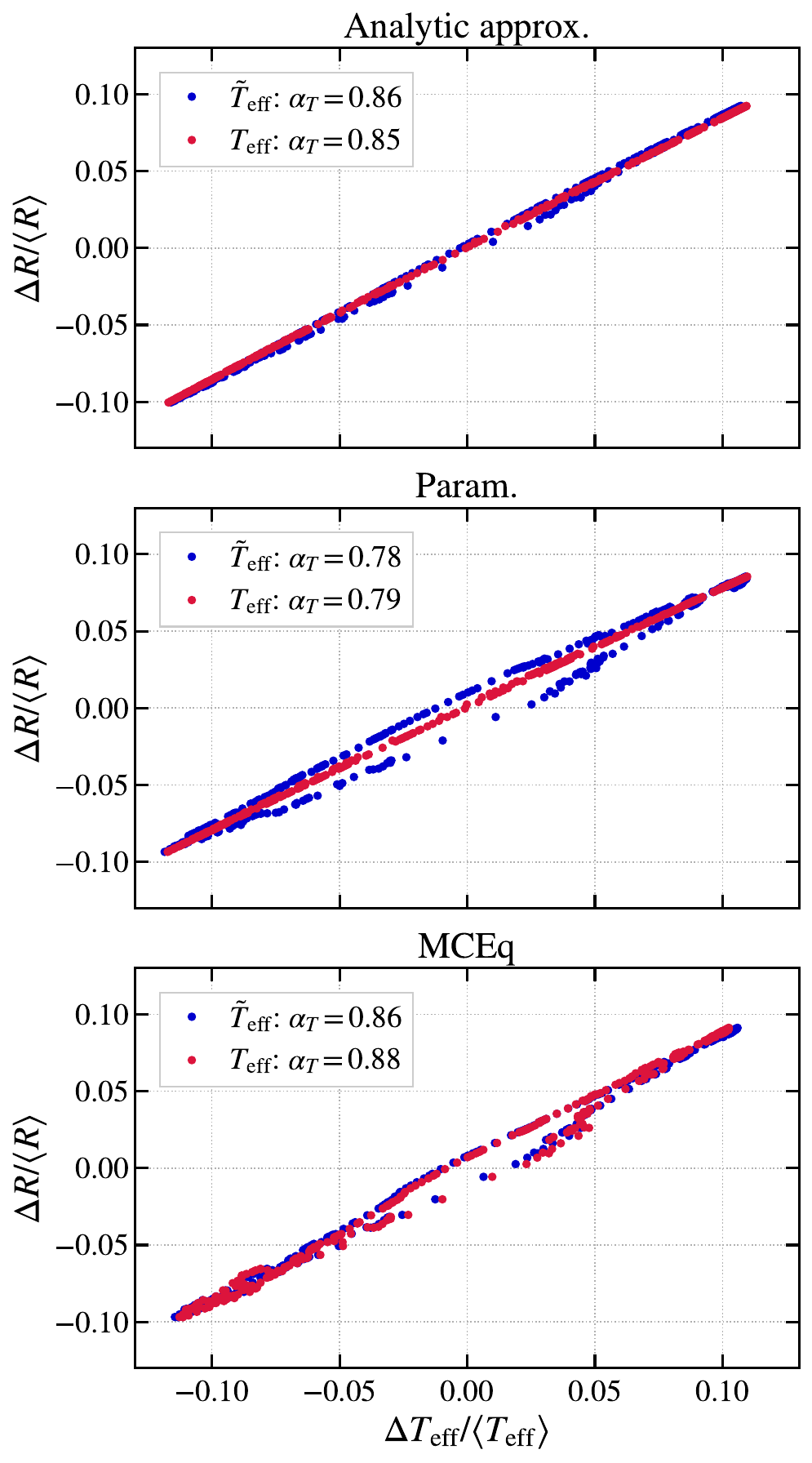}
    \vspace{-1.8em}
    
    \caption{Correlation between the relative variation in event rate and effective temperature calculated according to the methods from \refsecss{sec:analytic}{sec:mceq}{sec:param}. 
    Different colors indicate the 
    two definitions of effective temperature. Correlation coefficients calculated by fitting a line to the points are given in the legend.}
    \label{fig:alpha_comparison}
    \vspace{-1.2em}
\end{figure}

\begin{figure}
    \centering
    \includegraphics[width=\linewidth]{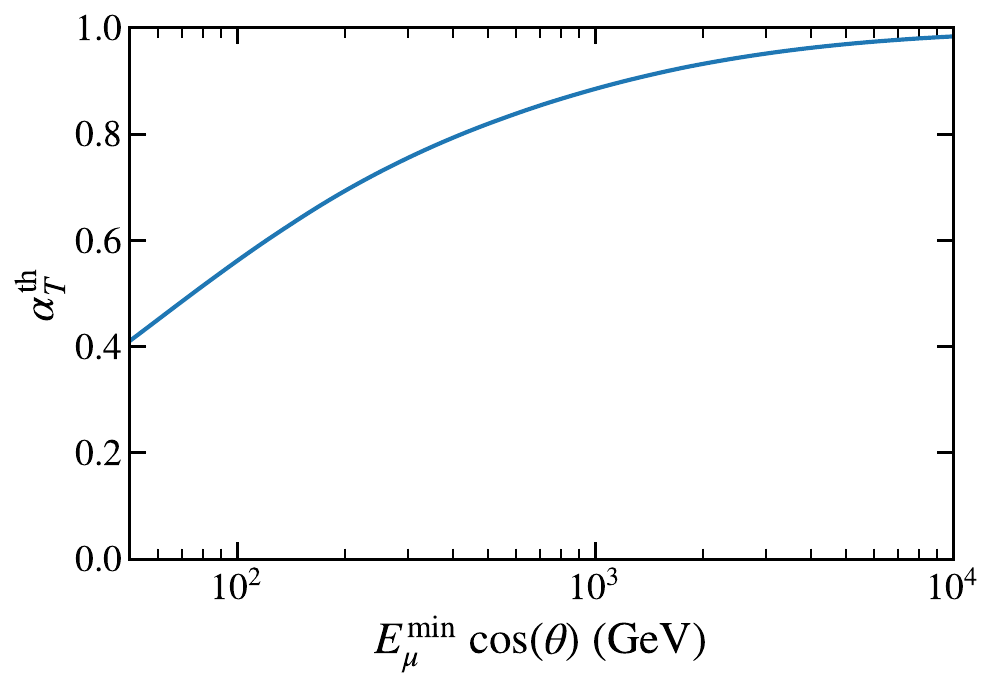}
    \caption{Theoretical prediction of the temperature correlation coefficient as a function of muon threshold energy and zenith angle, calculated using the analytic approximation of \refsec{sec:analytic} assuming an isothermal atmosphere with $T = \SI{220}{\kelvin}$.}
    \label{fig:alpha_theo}
\end{figure}

A theoretical expectation for the correlation coefficient as a function of zenith angle and threshold energy can be calculated by writing it in the following form:
\begin{equation}
\alpha_T^\mathrm{th}(\Emumin, \theta) = \frac{T}{I(\Emumin,\theta)}\frac{{\rm d}I(\Emumin,\theta)}{{\rm d}T}.
\label{eq:alpha_theo}
\end{equation}
Using the expression for the integral rate, \refeq{eq:compact}, together 
with the expression in \refeq{eq:integralPmu}, the theoretical correlation
coefficient for the integral muon spectrum can be estimated.  To do so, we assume
relatively small deviations  of $T(X)$ from $\langle T_{\rm eff}\rangle$. The result is 
shown for fixed $T = \SI{220}{\kelvin}$ in \reffig{fig:alpha_theo} as a function of 
$\Emumin\cos(\theta)$  (see also Eq.~\ref{eq:integralPmu}).  We limit the energy range at the lower end to \SI{50}{\giga\eV} as at lower energies muon decay is expected to have a non-negligible impact. At energies above \SI{10}{\tera\eV}, the muon prompt component is expected to become important, which will lower the value of \aT compared to the calculations including only contributions from pions and kaons~\cite{Desiati:2010wt}. A calculation of the theoretical \aT using the weights of Ref.~\cite{Grashorn:2009ey} is compared with a range of experimental results in Ref.~\cite{Agafonova:2018kce}. Calculation of the correlation coefficient for the differential case can be carried out equivalently, but is
less universal because it depends on the energy-dependent effective area, which is different
for each detector.

Experimental values of $\alpha_T$ are obtained by applying a linear fit to
${\Delta R} / \langle R\rangle$ as a function of 
${\Delta T_{\rm eff}}/\langle T_{\rm eff}\rangle$,
where $R$ and $T_{\rm eff}$ are the measured event rate and the corresponding calculated effective temperature (e.g. per day) and the denominators are the average
over the observation period (e.g. a year).  In \reffig{fig:alpha_comparison}, we show correlation plots with 
calculated rates for the hypothetical detector introduced in \refsec{sec:rate}. Values obtained for the correlation coefficients differ little between effective temperature definitions. A larger difference is present between the methods based on cascade equations and the muon profile parameterization method. The good agreement between the analytic approximation and the MCEq calculation has been shown before, including for the case of seasonal variations of neutrinos~\cite{Gaisser:2019xlw, IceCube:2023qem}. In \refref{Gaisser:2021bwj}, a comparison between the analytic approach and the parameterization suggests that the level of agreement between different calculations and experimental results depends on the energy range relevant to the detector. 
\section{Relative contributions from pions and kaons}

\begin{figure}
    \centering
\includegraphics[width=\linewidth]{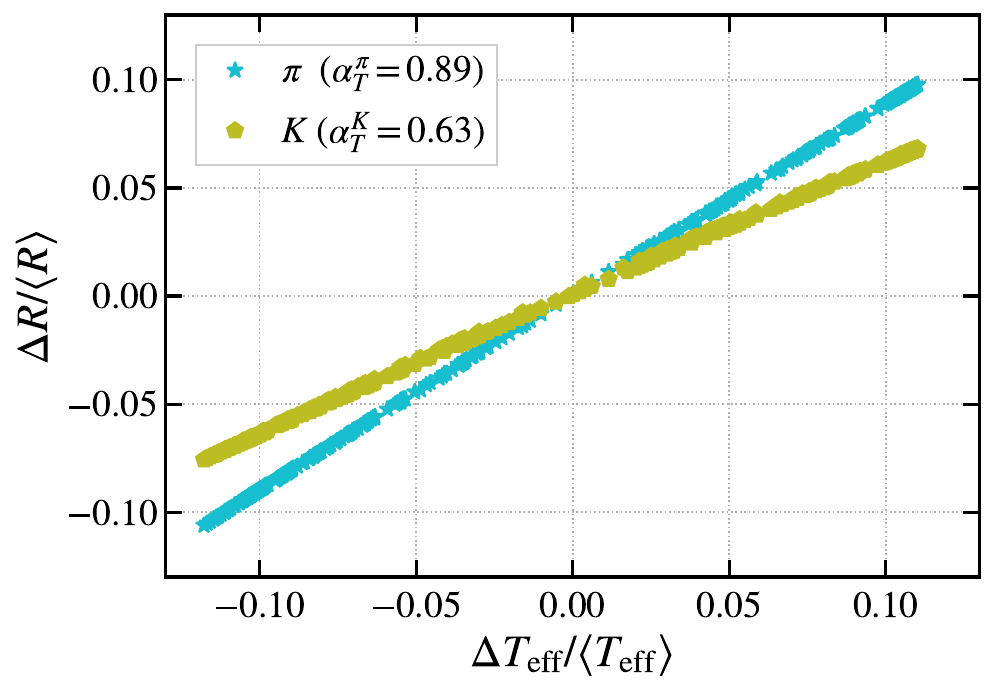}
    \caption{Variation of the muon rate originating from the pion and kaon channels plotted separately as a function of the effective temperature, calculated with the analytic approximation of \refsec{sec:analytic}.}
    \label{fig:alpha_K_pi_separate}
\end{figure}

The higher critical energy of kaons compared to pions results in a lower correlation with temperature for muons from kaon decay. This is illustrated in \reffig{fig:alpha_K_pi_separate}, where \aT is determined separately for the kaon and pion component in the calculation, $R = R_\pi + R_K$, using the analytic approximation \refeq{eq:Panalytic}. As a result, the measured correlation coefficient depends on the relative contribution of pions and kaons to the production of muons. A measurement of the seasonal variations of the atmospheric muon rate is therefore a probe of the atmospheric kaon-to-pion production ratio \rKpi.

\begin{figure}[tb]
    \centering
    \includegraphics[width=\linewidth]{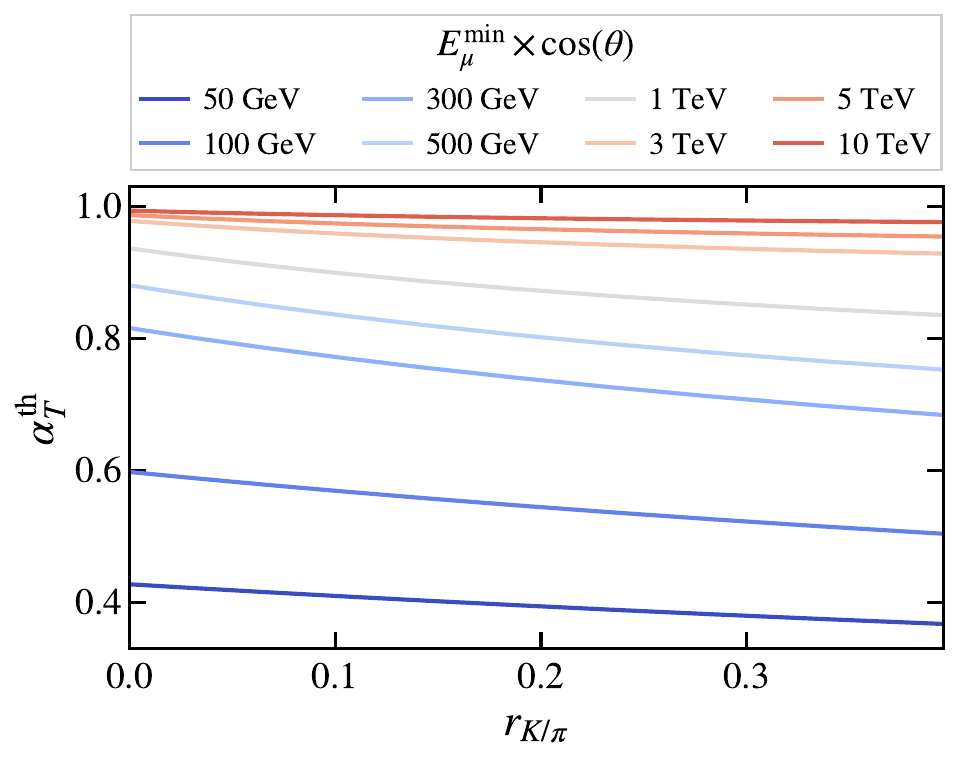}
    \caption{Theoretical expectations of the relation between \aT and the kaon-to-pion ratio $r_{K/\pi}$ calculated with the analytical approximation of the production spectrum given by \refeq{eq:integralPmu}. Spectrum weighted moments were assumed to be independent of energy.}
    \label{fig:alpha_Kpi_theo}
\end{figure}

In \refref{Grashorn:2009ey}, \rKpi was defined in terms of the spectrum weighted moments $Z_{NK}$ and $Z_{N\pi}$ as
\begin{equation}\label{eq:Kpi}
    r_{K/\pi} = \frac{Z_{NK}}{Z_{N\pi}}.
\end{equation}
The dependence of the correlation coefficient on the $K/\pi$ ratio can be estimated straightforwardly from the analytic approximation of \refsec{sec:analytic}, as the dependence on the spectrum weighted moments $Z_{N\pi}$ and $Z_{NK}$ is explicit in the parameters $A_{\pi\mu}$ and $A_{K\mu}$ of \refeq{eq:Api}. In this case, the correlation depends only on the ratio of $Z_{NK}$ and $Z_{N\pi}$. \reffig{fig:alpha_Kpi_theo} shows the theoretical expectation for $\alpha_T^\mathrm{th}$ for different $\Emumin \cos(\theta)$ from \refeq{eq:alpha_theo}, calculated as a function of \rKpi assuming $Z_{NK}$ and $Z_{N\pi}$ to be independent of energy, as in \refeq{eq:Z}. The nominal value of $K/\pi$ ratio is in this case taken to be $r_{K/\pi} = 0.0109/0.066 = 0.165$.

\begin{figure}[tb]
    \centering
    \includegraphics[width=\linewidth]{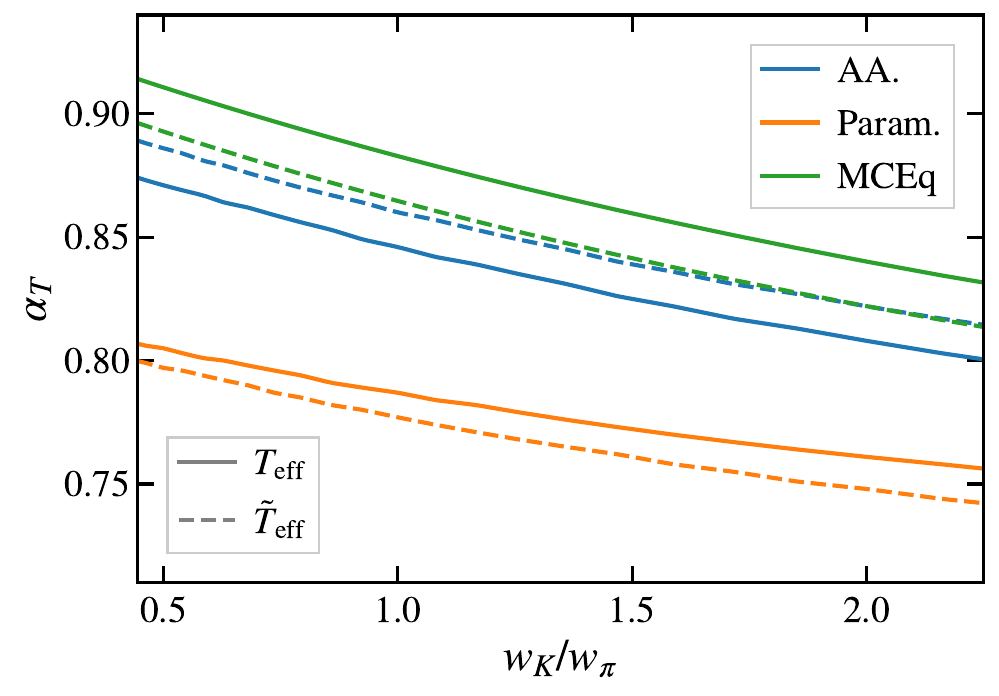}
    \caption{Relation between \aT and the ratio of $w_K$ and $w_\pi$, modifying the production of charged pions and kaons. The values are calculated for the detector of \reftab{tab:detector}, using the analytic approximation of \refsec{sec:analytic}, the parameterization of \refsec{sec:param}, and the MCEq calculation of \refsec{sec:mceq}. Both definitions of effective temperature given in \refsec{sec:Teff} are included.}
    \label{fig:alpha_Kpi}
\end{figure}

In \refref{Fedynitch:2019bbp}, a modified $K/\pi$ ratio was defined in terms of two weights $w_\pi$ and $w_K$ which scale the inclusive particle production spectrum,
\begin{equation}
    r^\star_{K/\pi} = \frac{Z_{NK}^\star}{Z_{N\pi}^\star} = \frac{w_K Z_{NK}}{W_\pi Z_{N\pi}} = \frac{w_K}{w_\pi} r_{K/\pi}.
\end{equation}
When using energy-dependent $Z$-factors or comparing different methods of calculating \aT, it is easier to express \aT as a function of $w_\pi$ and $w_K$ rather than the value of \rKpi itself. For calculations including only muons from the decay of $\pi^\pm$ and $K^\pm$, \aT will depend only on the ratio of the weights. In a full calculation including contributions from other channels, such as performed with MCEq, this simple relation breaks down. In \reffig{fig:alpha_Kpi}, a full calculation of of the expected \aT for the detector of \reftab{tab:detector} is shown as a function of $w_\pi/w_K$ for the analytic approximation, MCEq, and the parameterization. For the latter, the weights entered in the calculation of $f_\pi = (w_\pi Z_{N\pi}^{\gamma=1}) / (w_\pi Z_{N\pi}^{\gamma=1} + 0.635 w_K Z_{NK}^{\gamma=1})$, with $Z_{N\pi}^{\gamma=1}$ the energy-independent spectrum weighted moment for $\gamma=1$, as described in \refsec{sec:param}. For MCEq, the dependence was approximately estimated by scaling the production spectra of muons produced by pions and kaons with $w_\pi$ and $w_K$, respectively. The calculation of the effective temperatures and \aT is then repeated, as described in Section~\ref{sec:Teff}, with the scaled distributions.

\begin{figure}[tb]
    \centering
    \includegraphics[width=\linewidth]{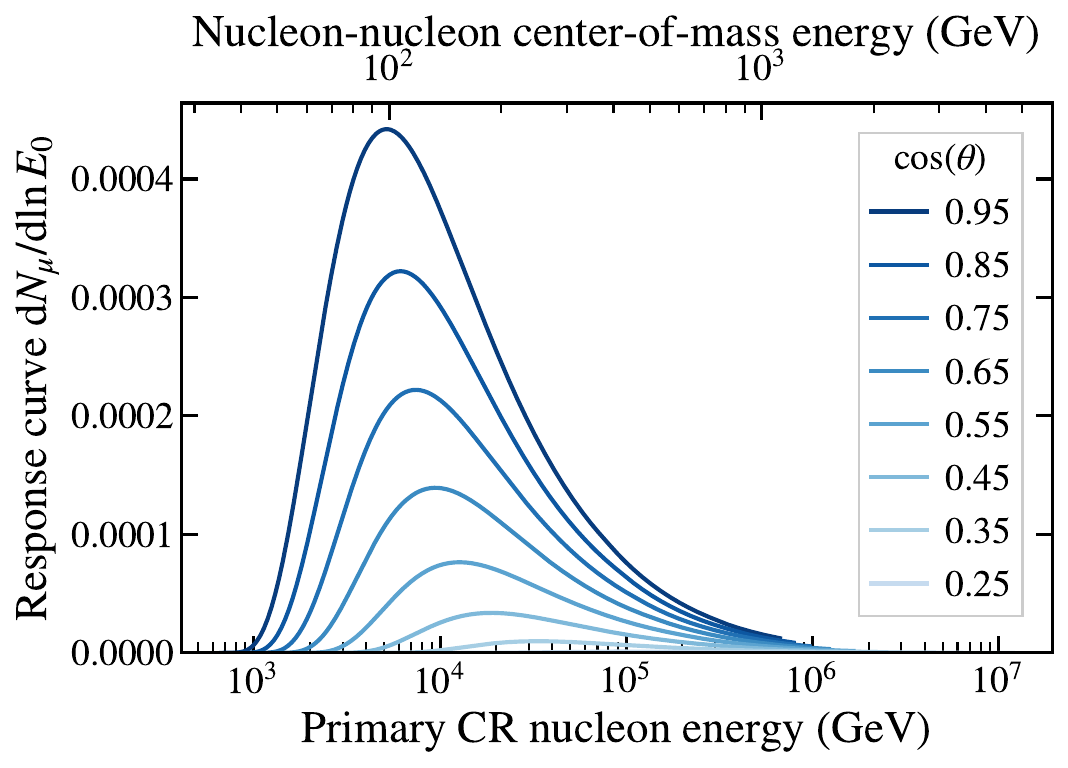}
    \caption{Response curve showing the contribution to the muon rate as a function of primary nucleon energy for the detector of \reftab{tab:detector}. Calculated using the parameterization of \refsec{sec:param}.}
    \label{fig:response_curve}
\end{figure}

Determining the experimental value of $\alpha_T$ is relatively insensitive to the assumed value of \rKpi, as the dependence in the calculation of the effective temperature mostly cancels out. By comparing the experimental result to the calculated correlation coefficient, it is possible to measure \rKpi for nucleon-nucleon interactions at median primary energies which are typically between 10-100 times the muon threshold energy at production, as illustrated in \reffig{fig:response_curve}. 

Preliminary results from IceCube were shown in \refref{Desiati:2011hea}. An alternative calculation of the relation between $\alpha_T$ and $r_{K/\pi}$ was shown earlier in \refref{Grashorn:2009ey} and used by other experiments such as Borexino~\cite{Borexino:2018pev}.

\section{Multiple muon events and nuclear primaries}\label{sec:multimu}

The traditional rate calculation as presented in \refsec{sec:rate} is based on the inclusive atmospheric muon flux. A shortcoming is that it does not take into account that muons produced in the same shower arrive at the detector simultaneously. While the muons arriving in bundles contribute individually to the calculated muon intensity, in realistic detectors they will often be indistinguishable, making the event rate lower than what is predicted from the calculation of \refeq{eq:rtheta}.

An estimate of the effect can be obtained for compact detectors by modifying the calculations presented in \refsec{sec:param}. Combining \refeqss{eq:compact}{eq:PofEmu}{eq:params}, the traditional rate calculation can be written as
\begin{align}
    R &= \Aeff\,I(\Emumin) \nonumber \\
              &= \Aeff\int_{\Emumin}\dd E_0\, \phi_N(E_0)\,\langle N(\Emumin, E_0)\rangle,
\end{align}
where $\langle N \rangle$ is the mean number of muons with energy above \Emumin produced by a nucleon with energy $E_0$, and we omit the $\theta$-dependence for conciseness. Writing the average as \mbox{$\langle N \rangle = \sum_0^\infty np(n)$}, with $p(n)$ the probability for a nucleon to produce a bundle of $n$ muons, shows explicitly that multiple muons get accounted for separately in the calculation rather than as a single event. Replacing this by the probability to have at least one muon above threshold per primary nucleon gives the expected intensity of bundles of muons with one or more muons above \Emumin,
\begin{equation}\label{eq:multimu_nucleon}
    I_\mathrm{bundle} (\Emumin) = \int_{\Emumin} \dd E_0\, \phi_N(E_0)\,\sum_{n=1}^\infty p(n|\langle N\rangle).
\end{equation}
Assuming the multiplicity to follow a Poisson distribution\footnote{\refref{Forti:1990st} finds that the multiplicity is described better by a negative binomial distribution.}, the sum is given by $1 - e^{-\langle N \rangle}$.

\begin{figure}[tb]
    \centering
    \includegraphics[width=\linewidth]{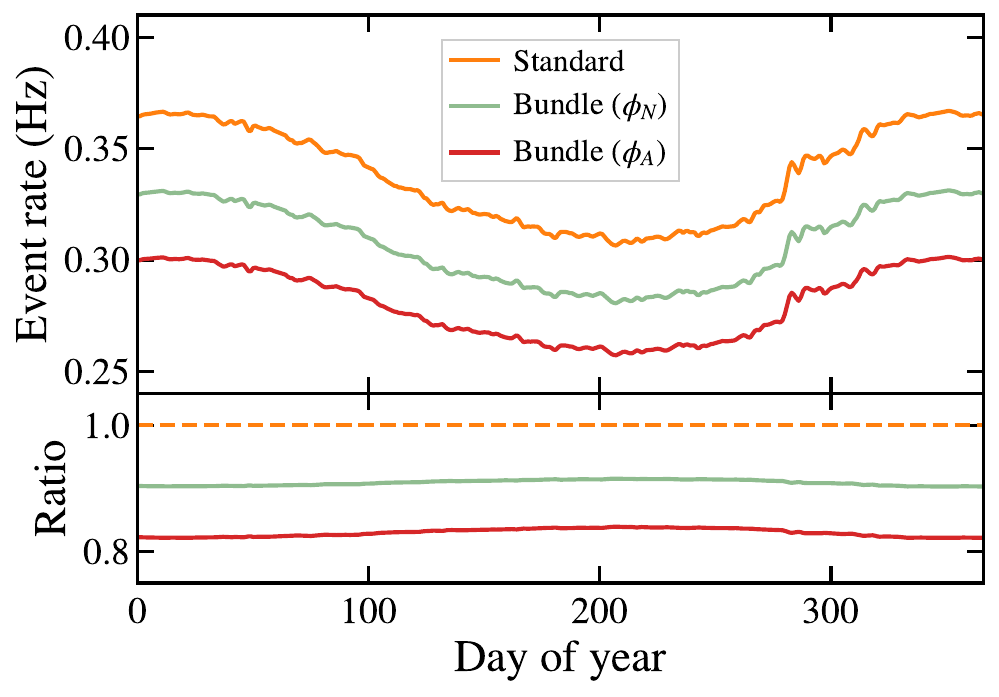}
    \caption{Daily event rate for the detector of \reftab{tab:detector} taking into account the effect of multiple muons from a single air shower arriving together. The standard calculation from the parameterization, already shown in \reffig{fig:rates}, is compared to the calculation of \refeq{eq:multimu_nucleon} using the total H3a nucleon flux, and of \refeq{eq:multimu_nucleus} using the mass composition given by H3a. The multiplicity was assumed to follow a Poisson distribution of mean $\langle N \rangle$ given by \refeq{eq:mean_N}.}
    \label{fig:R_multimu}
\end{figure}

Another effect which will decrease the event rate compared to \refeq{eq:rtheta} is the fact that a fraction of the primary nucleons arrive at the Earth bound in nuclei, which are more likely to produce higher multiplicity bundles of muons arriving simultaneously. To take this into account, we can integrate over a realistic flux model, such as the H3a model,
\begin{multline}\label{eq:multimu_nucleus}
    I_\mathrm{bundle} (\Emumin) = \\ \sum_A \int_{A\Emumin} \dd E_A\, \phi_A(E_A)\,\sum_{n=1}^\infty p\left(n|\langle N(\Emumin, E_A, A)\rangle\right),
\end{multline}
where $\phi_A$ is the differential flux of element $A$ and the sum runs over the different primary nuclei in the flux model. Note that we still assume that the energy in the nucleus is divided evenly over the $A$ nucleons. The expectation $\langle N \rangle$, which depends on the atmospheric temperature profile, can be estimated by integrating the parameterized production profile \refeq{eq:formula},
\begin{multline}\label{eq:mean_N}
    \langle N \rangle (\Emumin, E_A, A, \theta, X, T) = \\ \int \dd X \frac{\dd N}{\dd X} (\Emumin, E_A, A, \theta, X, T).
\end{multline}
The effect of taking muon multiplicity and a realistic nucleus flux into account is shown in \reffig{fig:R_multimu}. Performing the calculation using the total nucleon flux but taking into account multiple muon events decreases the expected rate by close to $10\%$. Taking into account also the mass composition of primary nuclei decreases the expectations by another 10\%.

\begin{figure}
    \centering
    \includegraphics[width=\linewidth]{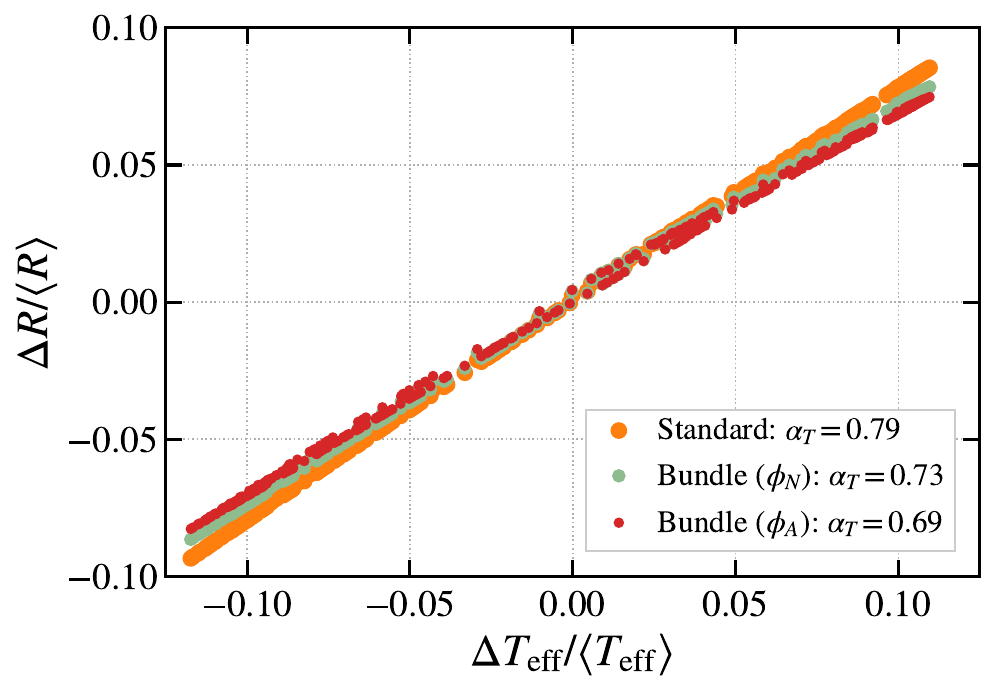}    \caption{Relative variations in the calculated event rate from \reffig{fig:R_multimu} as a function of variations in the effective temperature. Three rate calculations are shown -- Standard: each muon is individually counted in the rate; Bundle ($\phi_N$): muons produced by the same nucleon are counted as a single event; Bundle~($\phi_A$): muons produced by the same nucleus are counted as a single event. The effective temperature uses the standard calculation in each case.}
    \label{fig:alpha_multimu}
\end{figure}

It is of interest to examine how this modified rate calculation affects the expected correlation coefficient. The correlation plot including different rate calculations is shown in \reffig{fig:alpha_multimu}. Here, the effective temperature is taken to be the same in all cases, i.e. it is given by \refeq{eq:Teff}. This shows how the standard approach of comparing measured rates to the calculated \Teff may cause an underestimation of \aT. This may in turn lead to inaccuracies in the determination of $r_{K/\pi}$.

We note that this is a simplified estimate of this effect. A more accurate calculation can be obtained replacing the parameterized muon production profiles by production profiles obtained by using MCEq to solve the cascade equations for individual primary nuclei, or by performing a full simulation of the problem. This is especially important for geometrically extended detectors, where the energy threshold region needs to be treated in more detail.
\section{Summary}

The flux of atmospheric leptons varies with the seasons as the atmosphere contracts and expands, which influences the decay probability of parent mesons. The observation of this variation in the muon rate of underground detectors has a long history, and is usually analyzed in terms of its correlation with an effective temperature which is a weighted average of the atmospheric temperature profile. The magnitude of the correlation is expressed in terms of a correlation coefficient, which is sensitive to properties of the hadronic interactions in the atmosphere, specifically the kaon-to-pion production ratio.

The expected rate of muons can be calculated by integrating over the muon production spectrum multiplied by the effective area of the detector. An important difference exists between large-volume detectors where the effective area is energy dependent, and compact detectors at large depth, which can be approximated as energy independent (except for the dependence of the muon energy threshold on the zenith angle). Various approximations for calculating the muon production have been presented in the literature, each with their own advantages and disadvantages. We have considered here an approximate analytical solution to the atmospheric cascade equations, a code which numerically solves the cascade equations, and an approach where one integrates the muon production spectrum in individual air showers over the primary flux. Furthermore, different definitions of effective temperature have been used in the literature. A so-called derivative definition of the effective temperature, \refeq{eq:Teff}, follows naturally from the formalism, but is less straightforward to calculate than the simple average of the atmospheric temperature profile weighted by the muon production spectrum which has alternatively been used. In this work, we have compared several of these different methods and definitions, and showed how they lead to different predictions of the correlation coefficient \aT. We have also demonstrated the relation between the \aT and the kaon-to-pion production ratio. Finally, the relevance of multiple muon events and nuclear primaries was discussed, which are both not taken into account in the standard approach of an inclusive flux calculation from the total primary nucleon flux.

The treatment of seasonal variations of the atmospheric neutrino flux was not treated explicitly in this paper but can be carried out equivalently. With neutrinos originating dominantly from kaon decay above several hundred GeV, the temperature correlation is expected to be smaller compared to muons up to energies of several TeV~\cite{Desiati:2010wt, Tilav:2019xmf}. The different kinematics of neutrino production in the atmosphere thus make it possible to probe the K/pi ratio in an independent way using the same observatory. The feasibility has been demonstrated by the recent observation of seasonal variations of atmospheric neutrinos by IceCube~\cite{IceCube:2023qem}.

\vspace{1em}
\noindent {\bf Acknowledgments}: 
We thank our colleagues A.~Fedynitch, S.~Tilav, and D.~Seckel for helpful discussions related to this work. SV acknowledges funding from the National Science Foundation (NSF) grant \#2209483.

\appendix

\section{Cascade equations and their approximate solutions}\label{sec:appendix}

A simplified form of the cascade equation for the inclusive spectrum of charged pions 
in the atmosphere $\Pi(E,X)$ is~\cite{Gaisser:2016uoy}
\begin{align}
\frac{{\rm d}\Pi}{ {\rm d}X}=
&-\Pi (E,X){\left(\frac{1}{\Lambda_\pi}+\frac{\epsilon_\pi}{
  E\,X\,\cos(\theta)}\right)}\\&+\frac{Z_{N\pi}}{\lambda_N}\phi_N(E)\,e^{-X/\Lambda_N},
\label{eq:piapprox}
\end{align}
with a similar equation for the charged kaon channel.
The equation has two loss terms.  The first is from pion interactions in the atmosphere
where $\Lambda_\pi>\lambda_{\pi,{\rm int}}$ is an attenuation length for pions that accounts for 
their regeneration.  The second is the pion decay term, which depends on temperature, as in 
\refeq{eq:epsilon_pi}.  $X$ is the atmospheric slant depth along a direction with zenith angle $\theta$,
and the solution applies to a boundary condition at the top of the atmosphere
where $\Pi(E,X=0) = 0$ and $\phi_N(E)$ is the spectrum of nucleons evaluated at the energy
of the pion.  This form holds for a power-law spectrum of primary nucleons and for
production cross sections that depend only on the ratio $x_L$ of the lab energy of the secondary
particle to that of the parent.  In this case, the energy-dependence of the production
of the secondary is represented by the spectrum weighted moment, which for charged
pions is
\begin{equation}
Z_{N\pi}=\int_0^1(x_L)^{\gamma-1}F_{N\pi}(x_L)dx_L,
\label{eq:ZNpi}
\end{equation}
with $F_{N\pi}(x_L=E_\pi/E_N)$ the dimensionless inclusive particle production spectrum 
\begin{equation}
F_{N\pi} = \frac{E_\pi}{\sigma_{N,{\rm air}}}\frac{{\rm d}\sigma_{N,{\rm air}\rightarrow\pi}}{{\rm d}E_\pi}=E_\pi\frac{{\rm d}n_\pi(E_\pi,E_N)}{{\rm d}E_\pi},
\label{eq:sigma_inc}.
\end{equation}
which follows from the inclusive cross section $\sigma_{N,{\rm air}\rightarrow\pi}$, where $\sigma_{N,{\rm air}}$ is the inelastic nucleon-air cross section.

In application of this approximation, it is important to include
all intermediate channels in the calculation of the spectrum weighted
moments.  Especially important, for example, 
is $p+{\rm air}\rightarrow \Lambda + K^+ + xxx$, which has an important
influence on the muon charge ratio and on the energy dependence
of the kaon channel in general~\cite{Gaisser:2012zz}. Comparison~\cite{Gaisser:2019xlw}
of the approach given here with MCEq~\cite{Fedynitch:2015zma} that includes all 
intermediate channels shows only small differences, see also \refsecs{sec:rate}{sec:alpha}.
Generalizations
to include non-scaling behavior of the production cross sections
and energy-dependence of the primary spectral index are possible~\cite{Gondolo:1995fq,Gaisser:2019abc}.
However, the main justification for this simple approach, some version of
which has been used by many experiments, is that the seasonal variation
is itself a ratio in which many uncertainties cancel.

The solution of Eq.~\ref{eq:piapprox} for charged pions is
\begin{equation}
\Pi(E,X)=e^{-(X/\Lambda_\pi)}\frac{Z_{N\pi}}{\lambda_N}\phi_N(E)
           \int_0^X\exp\left [{-\frac{X'}{\Lambda^*_\pi}}\right ]
\left (\frac{X'}{ X}\right )^\frac{\epsilon_\pi}{E\cos(\theta)}\,
{\rm d}X',
\label{eq:exact}
\end{equation}
with $\Lambda^*_\pi = \Lambda_\pi\Lambda_N/(\Lambda_\pi-\Lambda_N)$.
In the high-energy limit, the
scaling limit solution of \refeq{eq:exact}, subject to the boundary condition
$\Pi (E,0)=0$, is
\begin{equation}
\Pi(E,X) \stackrel{E\gg\epsilon_\pi}{\longrightarrow} \phi_N(E, 0)\, \frac{Z_{N\pi}}{ 1-Z_{NN}}\, \frac{\Lambda_\pi}{\Lambda_\pi
-\Lambda_N}\, {\left( e^{-X/\Lambda_\pi}-e^{-X/\Lambda_N}\right) }.
\label{piofX}
\end{equation}
In the low energy limit,
\begin{equation}
\Pi (E,X)\stackrel{E\ll\epsilon_\pi}{\longrightarrow} \frac{Z_{N\pi}}{\lambda_N}
\phi_N(E, 0)\,e^{-X/\Lambda_N}\,\frac{X\,E\,\cos(\theta)}{\epsilon_\pi}.
\label{lowE}
\end{equation}

\begin{table}[b]

\centering
\setlength{\tabcolsep}{4pt}
\footnotesize
\begin{tabular}{lr}
Symbol & Value \\
\hline
$\gamma$ & 1.7 \\
$Z_{NN}$ & 0.262 \\
$Z_{N\pi}$ & 0.066 \\
$Z_{NK}$ & 0.0109 \\
$\lambda_N$ & \SI{85}{\g\per\cm\squared} \\
$\Lambda_N$ & \SI{115}{\g\per\cm\squared} \\
$\Lambda_\pi$ & \SI{148}{\g\per\cm\squared} \\
$\Lambda_K$ & \SI{147}{\g\per\cm\squared} \\
$r_\pi$ &  0.5731 \\
$r_K$ & 0.0458 \\
\hline
\end{tabular}
\caption{Constants used in the calculations, from \refref{Gaisser:2016uoy} (based on Sibyll 2.3~\cite{Riehn:2015oba}).}
\label{tab:constants}
\end{table}

Accounting for the two-body decay kinematics of $\pi^\pm\rightarrow \mu\,\nu_\mu$
leads to the muon production spectrum as an integral over the meson fluxes:
\begin{equation}\label{muprod}
\begin{split}
P_\mu(E_\mu,X) = \frac{\epsilon_\pi}{ X\cos(\theta)(1-r_\pi)}\,
                       &\int_{E_\mu}^\frac{\Emu}{r_\pi}
			\frac{\Pi(E,X)}{ E}\frac{{\rm d}E}{ E}  \\
                  + \frac{0.635\,\epsilon_K}{ X\cos(\theta)(1-r_K)}\,
		       &\int_{E_\mu}^\frac{\Emu}{r_K}
			\frac{K(E,X)}{ E}\frac{{\rm d}E}{ E}.
\end{split}
\end{equation}
Inserting the low- and high-energy limiting approximations for $\Pi(E,X)$ and
$K(E,X)$ into Eq.~\ref{muprod} leads to the corresponding expressions for
the low- and high-energy muon production spectra in \refeq{eq:mu-prod-low} 
and \refeq{eq:mu-prod-high}.

To check the accuracy of the approximation of \refeq{eq:low-high}, one can expand the exponentials in \refeq{eq:exact} and integrate to obtain
\begin{multline}
\Pi(E,X)=e^{-(X/\Lambda_\pi)}\frac{Z_{N\pi}}{\lambda_N}\phi_N(E) X\\
\times \left[\frac{1}{\alpha_\pi + 1}-\left(\frac{X}{\Lambda^*_\pi}\right)\frac{1}{\alpha_\pi + 2}+\frac{1}{2!}\left(\frac{X}{\Lambda^*_\pi}\right)^2\frac{1}{\alpha_\pi + 3}\cdots\right],
\label{eq:expand}
\end{multline}
where $\alpha_\pi=\epsilon_\pi/(E\cos(\theta))$.  Inserting this expression into
Eq.~\ref{muprod} and defining $z = E/E_\mu$ and $\xi_\pi = \epsilon_\pi/(E_\mu \cos \theta)$ then leads to a rapidly converging expression for the muon
production spectrum that can be evaluated numerically to compare with the
approximation of Eq.~\ref{eq:low-high}.  The series is
\begin{multline}
\label{eqnarrayPmu}
P_{\mu,\pi}(E_\mu,X) = \frac{e^{-X/\Lambda_\pi}}{1-r_\pi}\frac{Z_{N\pi}}{\lambda_N}\phi_N(E_\mu)\,\xi_\pi\\
\times\int_1^{\frac{1}{r_\pi}}\frac{{\rm d}z}{z^{\gamma+2}}\left[\frac{1}{z+\xi_\pi}-\frac{X}{\Lambda_\pi^*}\frac{1}{2z+\xi_\pi}+\frac{1}{2!}\left(\frac{X}{\Lambda_\pi^*}\right)^2\frac{1}{3z+\xi_\pi}...\right].
\end{multline}

\begin{figure}[tb]
    \vspace{0.8em}
    \centering
    \includegraphics[width=\linewidth]{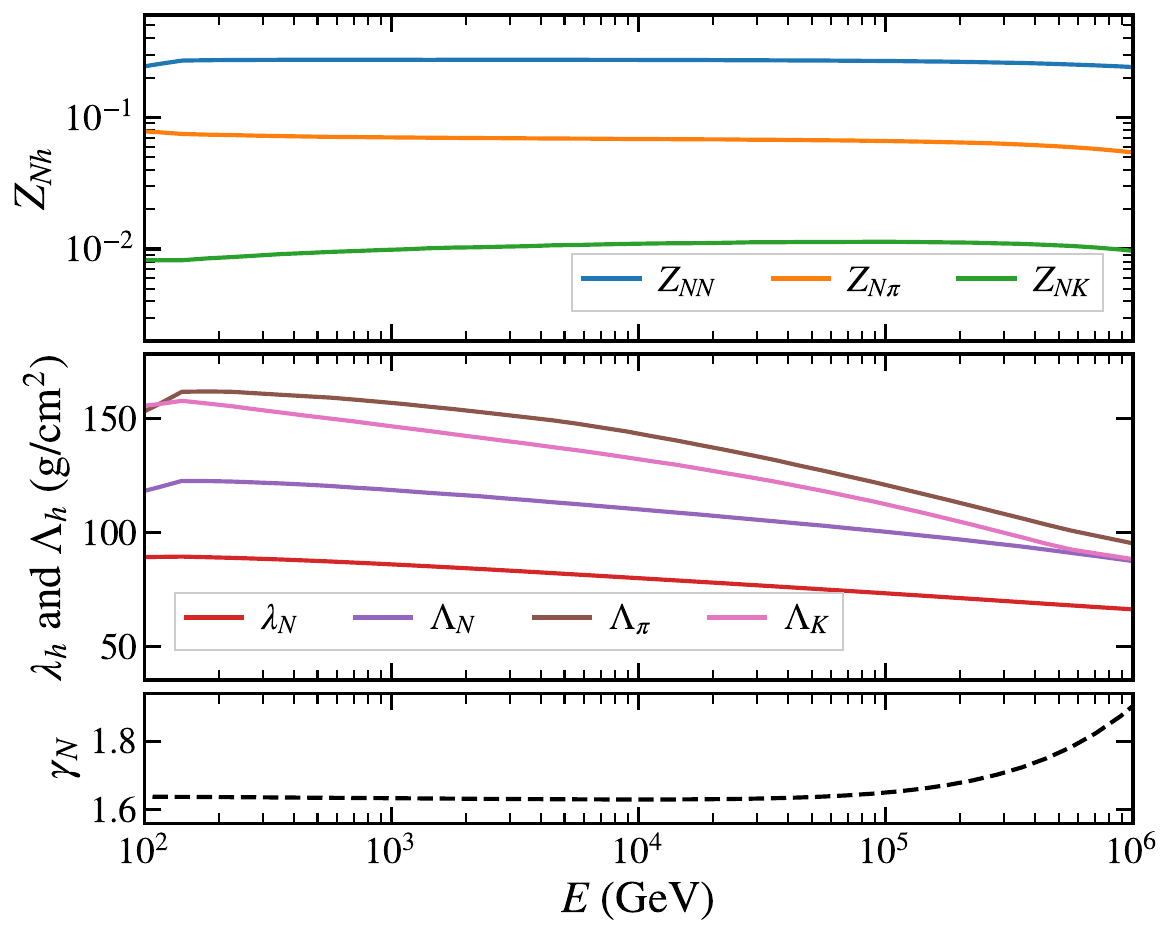}
    \caption{Energy dependent $Z$-factors, interaction lengths $\lambda$, and attenuation lengths $\Lambda$ as obtained from MCEq using H3a and Sibyll 2.3c.}
    \label{fig:not-constants}
\end{figure}

\begin{figure}[tb]
    \centering
    \includegraphics[width=\linewidth]{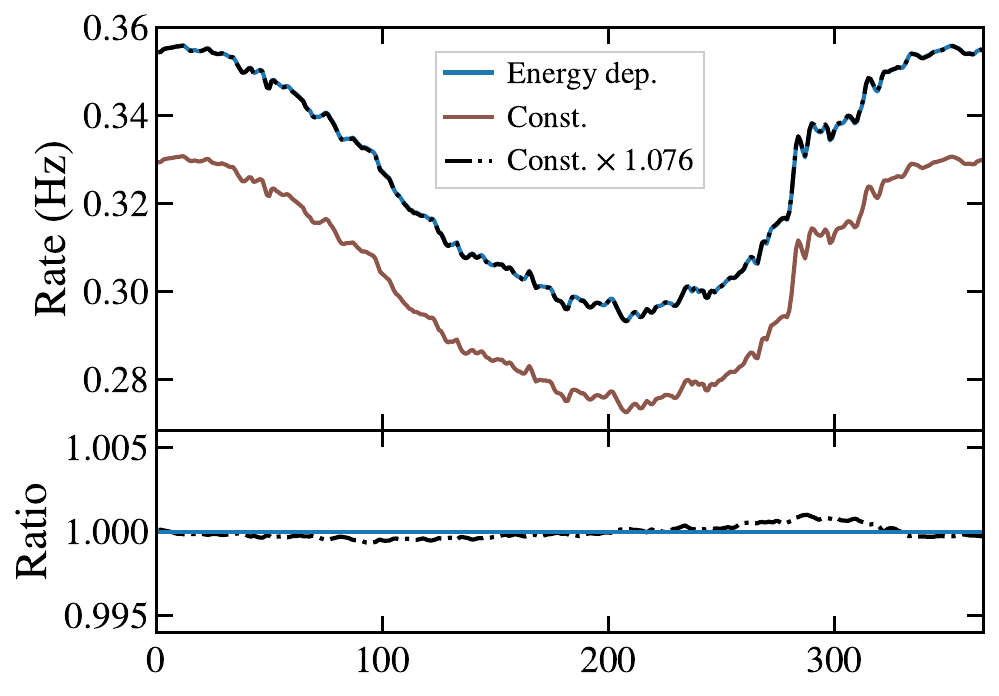}
    \caption{Rate calculated using the analytic approximation of the cascade equations. A comparison is made between the calculation using the constants given in \reftab{tab:constants}, and the energy-dependent values shown in \reffig{fig:not-constants}. The ratio of the two calculations has only a weak seasonal dependence.}
    \label{fig:const_vs_Edep}
\end{figure}

The constants used in this work are those relevant for \mbox{$E_\mu \sim \SI{1}{\tera\eV}$} from \refref{Gaisser:2016uoy}, repeated in \reftab{tab:constants}. The non-scaling cross sections and energy-dependent spectral index can be taken into account to first approximation by using energy dependent values for the parameters in the equations. Numerical values, shown in \reffig{fig:not-constants}, were obtained using MCEq and Sibyll 2.3c~\cite{Riehn:2019jet}. A comparison between the calculations using constants and energy-dependent parameters is shown in \reffig{fig:const_vs_Edep}. The difference in rate is nearly constant throughout the year, indicating that the energy-independent calculation is a valid approximation to determine the magnitude of the seasonal effect.

\bibliographystyle{elsarticle-num} 
\bibliography{references.bib}

\end{document}